\documentclass[preprint,authoryear,12pt]{elsarticle}
\usepackage{amsfonts,amssymb,amsmath,amsopn}
\usepackage{graphics}
\usepackage{rotating}
\usepackage{bigints}
\usepackage{booktabs}
\usepackage{subfigure}
\usepackage{url}
\usepackage{amssymb}
\usepackage{color}
\usepackage{multirow}
\usepackage{diagbox}
\usepackage{enumerate}
\usepackage[colorlinks=true]{hyperref}
\newcommand{\ignore}[1]{}
\newtheorem{asu}{{\sc Assumption}}

\newtheorem{thm}{Theorem}
\newtheorem{lem}{Lemma}
\newdefinition{rmk}{Remark}
\usepackage[margin=1in]{geometry}
\newproof{pf}{Proof}
\newproof{pot}{Proof of Theorem \ref{thm2}}

\usepackage{amssymb}
\usepackage{booktabs}

\allowdisplaybreaks
\def\dfrac{\displaystyle\frac}

\journal{\null}
\begin{document}
\begin{frontmatter}
\title{
Heteroscedasticity test of high-frequency data with jumps and microstructure noise
}
\author[NUS]{Qiang Liu}
\author[UM,UMZRI]{Zhi Liu}
\author[ZN]{Chuanhai Zhang}
\address[NUS]{Department of Mathematics, National University of Singapore, Singapore}
\address[UM]{Department of Mathematics, University of Macau, Macau SAR, China}
\address[UMZRI]{UMacau Zhuhai Research Institute, Zhuhai, China}
\address[ZN]{School of Finance, Zhongnan University of Economics and Law, Wuhan 430073, China}

\begin{abstract}
In this paper, we are interested in testing if the volatility process is constant or not during a given time span by using high-frequency data with the presence of jumps and microstructure noise.
Based on estimators of integrated volatility and spot volatility, we propose a nonparametric way to depict the discrepancy between local variation and global variation. 
We show that our proposed test estimator converges to a standard normal distribution if the volatility is constant, otherwise it diverges to infinity. 
Simulation studies verify the theoretical results and show a good finite sample performance of the test procedure. 
We also apply our test procedure to do the heteroscedasticity test for some real high-frequency financial data. 
We observe that in almost half of the days tested, the assumption of constant volatility within a day is violated. And this is due to that the stock prices during opening and closing periods are highly volatile and account for a relative large proportion of intraday variation. 
\end{abstract}
\end{frontmatter}

\section{Introduction}
It is well known that the logarithmic price of an asset is necessarily to be modeled as a semi-martingale process under the assumption of arbitrage-free and frictionless market.
The coefficient process driving the standard Brownian motion part, which is called the volatility process, serves as a measurement of risk in finance. 
Due to the wide applications of volatility in pricing of asset and derivative, portfolio selection, hedging and risk management, there are tons of research works on estimating the volatility. Many quantities targeting to measure the magnitude of the volatility, such as integrated volatility, spot volatility, realized Laplace transform of volatility, are proposed (see \citet{AJ2014} for their concrete definitions and a comprehensive introduction). 
Besides, it is also important to investigate the dynamic structure of the volatility process. 
Up until now, numerous models which have been proposed and widely applied are the ones in but not limited to \citet{BS1973},
\citet{V1977}, \citet{CIR1985}, \citet{C1992}, \citet{DH1993}. 
Or in another way, specific functional forms of the volatility process may be postulated before one can do goodness-of-fit tests to verify the correctness, related references are \citet{AS1996}, \citet{CW1999}, \citet{DLW2003},
\citet{DPV2006}, \citet{DP2008}, \citet{VD2012}, \citet{christensen2018diurnal}, and references therein.
Among them, one of the most basic questions have been tried to be answered is that whether the volatility process is constant or not over a period of time, say a day\footnote{Regarding the daily pattern of the volatility process, it reaches an agreement in \citet{TT1997,christensen2018diurnal,andersen2019time} that there are two distinct sources of variation for many financial asset return series. One of them is a deterministic diurnal component representing the fixed daily pattern. The other one is a stochastic part fluctuating around the fixed one, which brings in randomness and captures volatility clustering. Recently, \citet{christensen2018diurnal} concluded that the re-scaled log-returns are often close to homoscedastic within a trading day and  the fixed diurnal pattern accounts for a rather significant fraction of intraday variation in the volatility. But they also found that important sources of heteroscedasticity remain present in the data after annihilating the diurnal effect. Thus, the diurnal pattern is not sufficient to explain daily variation of the volatility.}. 
Putting forward a procedure to answer such a question is also the purpose of this paper. 
In the most of the previous literatures, the test procedures are
constructed based on a continuous diffusion assumption, while we consider the underlying data generating process of the
return as a general It$\hat{\text{o}}$ semi-martingale where the jumps are involved. 
Besides, the presence of market microstructure noise is also taken into account in our paper.
From a theoretical perspective, we contribute to propose a new nonparametric heteroscedasticity test procedure by using high-frequency data and further extend it to different settings incorporating the jumps and the market microstructure noise.

Our goodness-of-fit test procedure is based on the estimation of integrated volatility and spot volatility, which are well documented in existing literatures and many methods are valid under different settings. 
The integrated volatility quantifies the fluctuation of the asset price over a fixed time period, while the spot volatility measures the variation instantaneously.
We take a special case of diffusion process for an example to explain the mechanism implicated in our test. 
The stochastic process is discretely observed at evenly distributed points on the fixed time interval $[0,1]$. 
The asymptotic setup is of infilled type, namely, the mesh between the observation grids shrinks to zero. 
Under such a setting, we know that the estimators of the integrated volatility, for example, realized volatility and realized power variations (see e.g. \citet{ABDL2003}; \citet{BN2004}; \citet{BGJPS2006}; \citet{J2008}) are constructed based on all observations, while corresponding bounded kernel versions of the estimation of the spot volatility, as used in \citet{FW2008} and \citet{K2010}, only use the local data near a fixed time point. 
When the volatility remains constant over $[0,1]$, then the integrated volatility and the spot volatility at any given time second coincide, and the estimators of the former one have a faster convergence rate than the ones for the latter quantity. When the volatility varies over the time interval, the estimation of the spot volatility enables us to recover the time-varying volatility process, while the estimators of the integrated volatility give us a random variable. 
We construct a test statistic by integrating the squared differences between a sequence of spot volatility estimators over blocks with shrinking time length and an integrated volatility estimator. 
If the volatility is constant, then the scaled differences asymptotically distribute as a standard normal distribution, and the partial sum of the centered squared differences behaves asymptotically like a discrete martingale. 
Our statistic is shown to be asymptotic normal under the null hypothesis of constant volatility, while it diverges to infinity at an appropriate rate if the volatility process is time-varying. 
The test statistic is easy to compute and our test procedure can be naturally extended to other scenarios after taking the jumps and the market microstructure noise into consideration. 
Similar idea is also adopted in \citet{T2017} to test time-varying jump activity index for a pure jump semi-martingale defined on a fixed time interval.

We start our discussion as described above with continuous diffusion model, which is the most commonly-used one for the return process. 
But it has been shown, in \citet{BNS2006}, \citet{AJ2009}, \citet{AJ2010}, \citet{AJ2009deg}, \citet{JKLM2012} and references therein, that it is not adequate to describe
the various fluctuation patterns of financial asset price because of the presence of jumps, which may be due to the news shocks from the markets. 
The mixing jumps bring in extra bias compared with the estimation of volatility under the continuous framework. 
If the number of jumps is finite, two well-behaved estimators are realised multi-power variation estimator and realised threshold quadratic variation estimator, respectively. 
The former one was given in \citet{BN2004}, \citet{BSW2006} and \citet{J2008}, while the latter
one was proposed by \citet{M2009} and
\citet{MR2011}. 
The cases regarding more active jump intensity, like infinite activity or even infinite variation, are considered in \citet{JLK2014}, \citet{JT2014}, \citet{LLL2018} and among many others. 

Apart from jumps, the existence of market microstructure noise in the observation procedure, which may caused by the presence of a bid-ask spread and the corresponding bounces, the differences in trading sizes and in representation of the prices, the different informational content of price changes, the discreteness of price changes, and data errors, also brings in bias. 
To eliminate the bias, one of the most effective and easy to implement way is by averaging the raw data before we apply the aforementioned estimation procedure, which is called pre-averaging approach proposed in \citet{PV2009b} and further extended in \citet{JLMPV2009}. 
The other approaches are two time-scaled and multi time-scaled estimators proposed in \citet{ZMA2005} and \citet{Zhang2006}; the realised kernel method proposed in \citet{BNHLS2008a}; the quasi-maximum likelihood method in \citet{Xiu2010}, the local moment method proposed by \citet{bibinger2014estimating}, and etc.
Based on existing methods of volatility estimation in the presence of jumps and market microstructure noise, we extend our heteroscedasticity test procedure and verify our theoretical results.

The rest of this paper is organized as follows.
In Section \ref{sec2}, we give out our model setup and the asymptotic theoretical properties. We firstly illustrate our test under continuous semi-martingale assumption, and then extend our theories to the framework with jumps and market microstructure noise by using thresholding and pre-averaging techniques separately. 
In Section \ref{sec3}, we verify our theoretical results and test the finite sample performance of the proposed tests via Monte Carlo studies. Our tests are applied to some real high-frequency financial data sets for empirical
analysis in Section \ref{sec4}. In subsequent Section \ref{sec5}, we conclude our paper.
Technical proofs are postponed to Appendix.

\section{Theoretical results}\label{sec2}
In this section, we firstly construct our heteroscedasticity test procedure by modeling the logarithmic price process as a
continuous It$\hat{\text{o}}$ semi-martingale.
If a jump part of finite activity is further involved in the underlying data-generating process, the test procedure can be naturally extended after thresholding the raw observed data.
Finally, we incorporate the presence of market microstructure
noise into the observation procedure, namely the observed data at a given time equals to the value of the underlying process at that time plus another stochastic error term. 
We apply the pre-averaging technique before implementing the heteroscedasticity test to eliminate the bias due to the noise.
Detailed descriptions and assumptions regarding the jumps and the market microstructure noise will be given later.
By combining the techniques used for eliminating the effects of the jumps and the market microstructure noise, we can also extend the test procedure to the situation with simultaneous presence of the jumps and the market microstructure noise. Since the extension can be evidently seen from our previous results, we omit its detailed proof and discussion in this paper.

Throughout the paper, all the processes are defined on the time interval $[0,1]$. 
We denote $V_{i/n}$ to be the value of the
process $V$ at the time point $i/n$ and define $ \Delta_i^n V = V_{i/n} - V_{(i-1)/n} $ for $i=1,\dots, n$. 
The whole test procedure is based on an infill asymptotic setting, namely $n\rightarrow \infty$, which gives us the high-frequency data. 
We use the notations $\rightarrow^{p}, \rightarrow^{d}, \rightarrow^{ds}$ to denote convergence in probability, convergence in distribution and stable convergence, respectively. 
In general, we say $\mathcal{F}$-stable convergence of a sequence $X_n$ to $X$
defined on an extension of $(\Omega, \mathcal{F}, \mathcal{F}_t, P)$,  if for any bounded Lipschitz function $g$ and any bounded
$\mathcal{F}$-measurable $\mathcal{Q}$, as $n\rightarrow \infty $, it holds that
\begin{align*}
\mathbf{E}[\mathcal{Q}g(X_n)] \rightarrow \mathbf{E}'[\mathcal{Q}g(X)],
\end{align*}
where $\mathbf{E}'$ stands for the expectation on an extension space. 
The detailed definition and more properties of stable convergence can be found in \citet{JS2003}.

\subsection{Continuous semi-martingale}\label{part:cont}
At first, we present our methodology in a benchmark setup, which excludes jumps and market microstructure noise when modeling the high-frequency data.
We denote $X$ to be the logarithmic price process of an asset, and $X$ is set to be an one-dimensional continuous It$\hat{\text{o}}$ semi-martingale defined on the filtered probability space $(\Omega,\mathcal{F},(\mathcal{F}_{t})_{t\geq 0},
\mathcal{P})$ with the following form:
\begin{equation}\label{model1}
X_t= X_0 + \int_0^t b_sds+ \int_0^t \sigma_sdB_s,
\end{equation}
where $b$ and $\sigma$ are progressively measurable processes, and $B$ is a standard Wiener process. 
We also assume that the volatility process $\sigma$ to be a continuous It$\hat{\text{o}}$ semi-martingale on the same filtered probability space $(\Omega,\mathcal{F},(\mathcal{F}_{t})_{t\geq 0}, \mathcal{P})$, and it can be represented as
\begin{equation}
\begin{split}
\sigma_t = & \sigma_0 + \int_0^t b_s^{\sigma} ds + \int_0^t D_s^{\sigma} dB_s + \int_0^t D_s^{'\sigma} dB_s^{'},
\end{split}
\end{equation}
where $b^{\sigma}$, $D^{\sigma}$ and $D^{'\sigma}$ are adapted, c$\grave{\text{a}}$dl$\grave{\text{a}}$g stochastic processes, $b^{\sigma}$ is further predictable and locally bounded, and $B^{'}$ is another standard Wiener process independent of $B$. 
It is required that $\sigma$ is bounded away from 0, that is, $\sigma_t > 0$ for $0 \leq t \leq 1$ almost surely.
We note that the common driving standard Wiener process $B$ in $X$ and $\sigma$ accommodates the leverage effect in finance, which depicts the dependence structure between these two stochastic processes.
Such continuous semi-martingale models for the log-price processes and the volatility process are widely used in vast existing high-frequency literature for volatility estimation, e.g., \citet{BNHLS2008a}, \citet{MZ2009}, \citet{JLK2014}, and etc.

In this paper, we are interested in investigating the pattern of the volatility process. Specifically, we want to test if the volatility process is constant or not during a given time period. To this end, we partition the sample space $\Omega$
into two complementary subsets
\begin{align*}
\Omega^{c} = \{ \omega: \sigma_t(\omega) = \sigma_0(\omega), t \in [0,1] \}, \quad \Omega^{v} = \Omega \setminus \Omega^{c}.
\end{align*}
The null hypothesis can then be written as $\mathcal{H}_0: \omega \in \Omega^{c}$, while the alternative $\mathcal{H}_a: \omega \in \Omega^{v}$. 
Our target then turns to proposing a test with a pre-set asymptotic significance level and with power going to one to test the null hypothesis, as $n\rightarrow \infty$.

We start demonstrating our theories with the estimation of  integrated volatility
$IV := \int_0^1 \sigma_s^2 ds$. It is well known that the most frequently used estimator of the integrated volatility is the so-called realized volatility, which is defined as
\begin{equation}
\widehat{IV}^{n} = \sum_{i=1}^{n} (\Delta_i^n X)^2.
\end{equation}
It is shown in \citet{BNS2007} that, under our setting,
\begin{equation}\label{cltint}
 \dfrac{\sqrt{n}(\widehat{IV}^{n} - IV )}{\sqrt{\int_0^t2\sigma_s^4ds}}  \rightarrow^{d} N(0,1),
\end{equation}
where $N(0,1)$ denotes standard normal distribution with mean 0 and variance 1. And the central limit theorem result can be turned feasible when we replace the integrated quarticity $\int_0^t\sigma_s^4ds$ by its consistent estimators. 
Based on the estimation of the integrated volatility, the estimation of spot volatility $\sigma_{\tau}^2$ at any given time $\tau$ can be correspondingly proposed by applying the kernel method described in \citet{FW2008}, \citet{K2010}, \citet{YFLZZ2014} and \citet{LLL2018}. 
For example, using the specific one-side uniform kernel function $K(u) = 1_{ \{ 0
\leq u\leq 1 \} }$, we can obtain an estimator of $\sigma_{\tau}^2$ as
\begin{equation}
\widehat{\sigma^2}_{\tau}^{n}(k_n) = \dfrac{1}{k_n\Delta_n}\sum_{i=\lfloor \tau\cdot n  \rfloor + 1}^{\lfloor \tau \cdot n \rfloor +k_n }
(\Delta_i^n X)^2,
\end{equation}
where $k_n$ is the number of intervals after the time point $\tau$ and lie closest to $\tau$. 
Following the theoretical results in aforementioned references, we conclude that under our setting, and if further $k_n\rightarrow \infty$ and $k_n^2/n \rightarrow 0$
hold, as $n \rightarrow \infty$, we have
\begin{equation}\label{cltspo}
\sqrt{k_n} \dfrac{\widehat{\sigma^2}_\tau^{n}(k_n) - \sigma_{\tau}^2}{\sqrt{2\sigma^4_\tau}} \rightarrow^{d} N(0,1), ~~\mbox{as}~~ n \rightarrow \infty.
\end{equation}
The consistency result $(\widehat{\sigma^2}_\tau^{n}(k_n))^2 \rightarrow^{p} \sigma_{\tau}^4$ further implies that a feasible central limit theorem can be obtained if we replace $\sigma_{\tau}^4$ by $(\widehat{\sigma^2}_\tau^{n}(k_n))^2$.

Now, we state the first test procedure, which is based on the estimators of the integrated volatility and the spot volatility discussed above. 
\begin{thm}\label{thm1}
$X$ follows the process in \eqref{model1}. 
\begin{enumerate}
\item For $ \omega \in \Omega$, if as $n \rightarrow \infty$, $k_n \rightarrow \infty$ and $k_n/n \rightarrow 0$, then it holds that, as $n \rightarrow \infty$, 
	\begin{equation}\label{thm1_1}
	\dfrac{k_n}{n} \sum_{j=0}^{\lfloor n/k_n \rfloor-1} (\widehat{\sigma^2}_{jk_n/n}^{n}(k_n)-\widehat{IV}^{n} )^2 \rightarrow^{p} \int_{0}^{1} (\sigma_s^2 - IV)^2ds.
	\end{equation}
\item For $\omega \in \Omega^{c}$, if as $n \rightarrow \infty$, $k_n \rightarrow \infty$ and $k_n^2/n \rightarrow 0$, then it holds that, as $n \rightarrow \infty$, 
\begin{equation}\label{thm1_2}
\mathcal{T}^{n}(k_n):= \sqrt{\dfrac{k_n}{2n}} \sum_{j=0}^{\lfloor n/k_n \rfloor-1 } \Big\{ \big(\dfrac{\sqrt{k_n}(\widehat{\sigma^2}_{jk_n/n}^{n}(k_n)-\widehat{IV}^{n})}{\sqrt{2}\widehat{IV}^{n}}\big)^2 - 1\Big\} \rightarrow^{d} N(0,1).
\end{equation}
\item Denote $ z_{\alpha} $ as the $\alpha$-quantile of standard normal distribution, if as $n \rightarrow \infty$, $k_n \rightarrow \infty$
and $k_n^2/n \rightarrow 0$, then it holds that, as $n \rightarrow \infty$, 
\begin{align}\label{thm1_3}
\begin{cases}
&\mathcal{P}( \mathcal{T}^{n}(k_n) > z_{1-\alpha}|\Omega^{c}) \rightarrow \alpha, \ \text{if} \ \mathcal{P}(\Omega^{c}) > 0, \\
&\mathcal{P}( \mathcal{T}^{n}(k_n) > z_{1-\alpha}|\Omega^{v}) \rightarrow 1.
\end{cases}
\end{align}
\end{enumerate}
\end{thm}
%

The intuition is as follows. If the volatility process is constant over $[0,1]$, then the magnitudes of the integrated volatility and the spot volatility at any given time are equal. 
So that from the conclusion $(\ref{cltspo})$, it holds that $ \sqrt{k_n}(\widehat{\sigma^2}_{jk_n/n}^{n}(k_n) - IV) $ with $j=0,...,\lfloor n/k_n \rfloor-1$ are asymptotically uncorrelated and normally distributed with mean
zero and variance $2\sigma^4_{0}$. 
Then, by martingale central limit theorem in \citet{HH1980}, we can obtain
\begin{equation*}
\sqrt{\dfrac{k_n}{2n}} \sum_{j=0}^{\lfloor n/k_n \rfloor -1 } \Big( \big(\dfrac{\sqrt{k_n}(\widehat{\sigma^2}_{jk_n/n}^{n}(k_n)-IV)}{\sqrt{2}\sigma^2_{0}}\big)^2 - 1\Big) \rightarrow^{d} N(0,1), ~~\mbox{as}~~ n \rightarrow \infty.
\end{equation*}
After replacing $IV$ by its estimator $\widehat{IV}^{n}$ above, we note that the independence structure between the terms in the summation is broken up. 
But it does not affect the asymptotic conclusion, because $\widehat{IV}^{n}$ converges to $IV$ at a faster rate,
compared with the convergence rate of $(\widehat{\sigma^2}_{jk_n/n}^{n}(k_n) - \sigma^2_{jk_n/n})$ to zero. 
After substituting $\sigma^2_{0}$ with its consistent estimator $\widehat{IV}^{n}$, we obtain the result (\ref{thm1_2}). Whether the volatility process is constant or not, we always have (\ref{thm1_1}), where the left hand side term is approximately the Riemann sums of the term on the right hand side. 
Plugging (\ref{thm1_1}) into (\ref{thm1_2}), we see that if the constant volatility assumption is violated, then the quantity $\mathcal{T}^{n}(k_n)$ in (\ref{thm1_2}) will tend to infinity at a fast rate of $\sqrt{nk_n}$. 
The different asymptotic properties of $\mathcal{T}^{n}(k_n)$ for constant volatility and time-varying volatility lead to conclusion (\ref{thm1_3}) and enable us to do the hypothesis testing in our way.

\subsection{Finite activity jump}\label{part:jump}
Now, we consider the setting where the underlying logarithmic price process is modeled as the combination of the continuous process $X$ and another pure jump process $J$, which is restricted to be of finite activity. 
That is, we have
\begin{equation}\label{model2}
Y_t = X_t + J_t,
\end{equation}
and the process $Y$, instead of $X$ in the last part, is observed at the time points $\frac{i}{n}$, for $i=0,1,...,n$. We write $J_t = \sum_{j=1}^{N_t} \gamma_{\tau_j}$, where $N_t$ is a non-explosive counting process with possibly time varying intensity, $\gamma_{\tau_j}$ is the size of the jump at time $\tau_j$. These jump sizes are not necessarily i.i.d random variables, nor independent of $N$.

To eliminate the influence of jumps on estimating the integrated volatility, \citet{M2009} proposed a thresholding technique to discriminate the time intervals with jumps from those without jumps. 
Such a filtering procedure can be done by using a deterministic
threshold function $r(x)$ satisfying the following conditions:
\begin{asu}\label{asu2}
The function $r(x): \mathbf{R} \rightarrow \mathbf{R}$ satisfies, $\lim_{x\rightarrow 0}\dfrac{x\log(\dfrac{1}{x})}{r(x)} = 0$,
$\lim_{x\rightarrow 0} r(x) = 0$.
\end{asu}
It is shown that for the sample paths, with probability one, there are jumps between $[(i-1)/n, i/n]$ if $(\Delta_i^nY)^2 > r(1/n)$.
Since these intervals with jumps are finite, excluding observed data in these intervals has no influence on the asymptotic properties of the estimator of the integrated volatility. Consequently, the thresholding versions of the estimators of the integrated volatility (called truncated realised volatility) and the spot volatility are formalized as
\begin{equation}
\begin{split}
\widehat{IV}^{n,Thr} & = \sum_{i=1}^{n} (\Delta_i^n Y)^2 \mathbf{I}_{\{(\Delta_i^n Y)^2 \leq r(1/n) \}},\\
\widehat{\sigma^2_\tau}^{n,Thr}(k_n) & = \dfrac{1}{k_n\Delta_n}\sum_{i=\lfloor \tau\cdot n  \rfloor + 1}^{\lfloor \tau \cdot n
\rfloor +k_n} (\Delta_i^n Y)^2\mathbf{I}_{\{(\Delta_i^n Y)^2 \leq r(1/n) \}}.\\
\end{split}
\end{equation}
The same conclusions in (\ref{cltint}) and (\ref{cltspo}) also hold if we replace $\widehat{IV}^{n}$ and $\widehat{\sigma^2_\tau}^{n}(k_n)$
with $\widehat{IV}^{n,Thr}$ and $\widehat{\sigma^2_\tau}^{n,Thr}(k_n)$, respectively. As a by-product, their detailed proofs are also given as we prove the following main theorem in Appendix.

\begin{thm}\label{thm2}
$X$ follows the process in \eqref{model1}, and Assumption \ref{asu2} hold.
\begin{enumerate}
\item For $ \omega \in \Omega$, if $k_n \rightarrow \infty$ and $k_n/n \rightarrow 0$ hold, as $n \rightarrow \infty$, then we have, as $n \rightarrow \infty$, 
	\begin{equation}\label{thm2_1}
	\dfrac{k_n}{n} \sum_{j=0}^{\lfloor n/k_n \rfloor-1} (\widehat{\sigma^2}_{jk_n/n}^{n,Thr}(k_n)-\widehat{IV}^{n,Thr} )^2 \rightarrow^{p} \int_{0}^{1} (\sigma_s^2 - IV)^2ds.
	\end{equation}
\item For $\omega \in \Omega^{c}$, if $k_n \rightarrow \infty$, $k_n^2/n \rightarrow 0$ and $\sqrt{k_n}\log{n}/\sqrt{n} \rightarrow 0$ hold, as $n \rightarrow \infty$, then we have, as $n \rightarrow \infty$, 
	\begin{equation}\label{thm2_2}
	\mathcal{T}^{n,Thr}(k_n):= \sqrt{\dfrac{k_n}{2n}} \sum_{j=0}^{\lfloor n/k_n \rfloor -1 } \Big( \big(\dfrac{\sqrt{k_n}(\widehat{\sigma^2}_{jk_n/n}^{n,Thr}(k_n)-\widehat{IV}^{n,Thr})}{\sqrt{2}\widehat{IV}^{n,Thr}}\big)^2 - 1\Big) \rightarrow^{d} N(0,1).
	\end{equation}
\item Denote $ z_{\alpha} $ as the $\alpha$-quantile of standard normal distribution, if $k_n \rightarrow \infty$, $k_n^2/n \rightarrow 0$ and $\sqrt{k_n}\log{n}/\sqrt{n} \rightarrow 0$ hold, as $n \rightarrow \infty$, then we have, as $n \rightarrow \infty$, 
	\begin{align}
	\begin{cases}
	&\mathcal{P}( \mathcal{T}^{n,Thr}(k_n) > z_{1-\alpha}|\Omega^{c}) \rightarrow \alpha, \ \text{if} \ \mathcal{P}(\Omega^{c}) > 0, \\
	&\mathcal{P}( \mathcal{T}^{n,Thr}(k_n) > z_{1-\alpha}|\Omega^{v}) \rightarrow 1.
	\end{cases}
	\end{align}
\end{enumerate}
\end{thm}

%

To reduce the effect of jumps (finite activity or infinite activity) in estimating the integrated volatility, another alternative method is the so-called realized multi-power variation estimator (see \citet{BN2004}, \citet{BSW2006} and \citet{J2008}), which diminishes the effect of jumps by using the products of the consecutive absolute increments $|\Delta_{i}^n Y|$. 
Theoretically, both of these two estimators are rate-efficient, but the truncated realised volatility is more efficient than the realised multi-power variation estimator in the sense of having a smaller variance. 
Indeed, the realised multi-power variation estimator is mainly biased by large jumps but is less affected by small jumps, while on the contrary, the truncated realised volatility is problematic in removing small jumps but eliminates large jumps effectively. In \citet{Veraart2011}, the properties of these two estimators are analyzed and compared comprehensively, their finite sample performances are verified by numerous Monte Carlo studies under different models. 
Furthermore, a combination of these two estimators breeds a new estimator called truncated realized multi-power variation estimator therein, which achieves the best effect of finite sample performance, since such a combination compensates the weaknesses of these two estimators. We note that our test procedure can be constructed accordingly by using these estimators mentioned, but we only consider the truncated realised volatility version here from the perspective of both simplicity and efficiency.
	
\begin{rmk}
	The restriction of finite activity on the jump process $J$ can be relaxed to some extent, for example, the case of
	L\'{e}vy jumps of infinite activity with finite variation. 
	It can be shown that the same conclusions in above theorem also hold for this relatively relax condition, but we only consider finite jumps for simplicity of the proof procedure. More on related properties and analyses can be found in \citet{MR2011} and \citet{JLK2014}.
\end{rmk}
\begin{rmk}
	One possible choice for $r(x)$ is the power function $cx^{\omega}$, with $c$ being a constant and $\omega \in (0,1)$. A time
	varying version of $r(x)$ (may be stochastic) is considered in \citet{MR2011}. Furthermore, \citet{AJ2009} point out that the value
	of $c$ should be proportional to the ``average" value of $\sigma_t$, which could be consistently estimated by the multi-power variation estimator mentioned above. The specific setting of the parameters $c$ and $\omega$ are also discussed
	in \citet{Veraart2011}, supported by a great deal of simulation studies.
\end{rmk}

\subsection{Market microstructure noise}\label{part:noise}
In this part, the data generating process of log-price is still modeled as the continuous semi-martingale $X$, but the observation procedure is conducted with disturbance. 
Mathematically, the observed data $Z_{i/n}$ at $\frac{i}{n}$ for $i=0,1,...,n$ are the underlying process
$X_{i/n}$ contaminated by another market microstructure noise term $\epsilon_{i/n}$, that is
\begin{equation}\label{model3}
Z_{i/n} = X_{i/n} + \epsilon_{i/n}.
\end{equation}
For the convenience of description, we define $\epsilon_t$ over the whole time span for $t \in [0,1]$. About the process $\epsilon$,
we assume that there exists a transition probability $Q_t(\omega,dx) $ from $(\Omega,\mathcal{F}_t) $ into $R$. We endow the space
$\Omega' = R^{[0,\infty)}$ with the product Borel $\sigma$-field $\mathcal{F}'$ and with the probability $\mathcal{Q}(\omega,d\omega')$
which is the product $\otimes_{t \geq 0} Q_t(\omega,\cdot)$. The process $Z$ is called the ``canonical process" on $(\Omega', \mathcal{F}')$,
with the filtration $\mathcal{F}' = \sigma(Z_s: s\leq t)$. We then work in the filtered probability space $(\Omega'', \mathcal{F}^{''},
\mathcal{F}_{t \geq 0}^{''}, \mathcal{P})$ with
\begin{equation*}
\Omega'' = \Omega \times \Omega', \mathcal{F}^{''} = \mathcal{F} \times \mathcal{F}^{'}, \mathcal{F}_t = \cap_{s>t}\mathcal{F}_s \times
\mathcal{F}_s^{'}, \mathcal{P}^{''}(d\omega,d\omega^{'}) = \mathcal{P}(d\omega)\mathcal{Q}(\omega,d\omega').
\end{equation*}
And the following assumption is satisfied:
\begin{asu}\label{asu3}
We have
\begin{equation*}
\int x Q_t(\omega,dx) = X_t(\omega),
\end{equation*}
and the process
\begin{equation*}
\alpha_t(\omega) = \int x^2 Q_t(\omega,dx) - X_t(\omega)^2 = \mathbf{E}[(Z_t)^2|\mathcal{F}](\omega) - X_t(\omega)^2
\end{equation*}
is c$\grave{a}$dl$\grave{a}$g(necessarily ($\mathcal{F}_t$)- adapted), and the process
\begin{equation*}
\beta_t(\omega) = \int x^8Q_t(\omega,dx)
\end{equation*}
is locally bounded.
\end{asu}

Before giving our estimators of the integrated volatility and the spot volatility, we firstly need to pre-average the raw increments with
a function $g$ supported on the interval $[0,1]$ satisfying
\begin{asu}\label{asu4}
The function $g$ is continuous and piecewise  differentiable with a piecewise Lipschitz derivative $g'$,
\begin{equation*}
g(0) = g(1) = 0, \qquad 0 < \int_{0}^{1} g(s)^2ds < \infty.
 \end{equation*}
\end{asu}
Denote the shorthand $g_i^n = g(i/p_n)$, then the pre-averaged increments for any process $V$ is defined as
\begin{equation*}
\overline{V}_{jp_n}^n = \sum_{i=1}^{p_n} g_i^n \Delta_{jp_n + i}^nV, \qquad \text{for} \quad j = 0,\cdots, \lfloor n/p_n \rfloor-1.
\end{equation*}
Now, these treated increments are used to construct our estimator of the integrated volatility, which is given by
\begin{equation}
\widehat{IV}^{n,Pre}(p_n) = \dfrac{1}{\varphi_n} \sum_{j=0}^{ \lfloor n/p_n \rfloor-1}(\overline{Z}_{jp_n}^n)^2,
\end{equation}
with $\varphi_n = \dfrac{1}{p_n}\sum_{i=1}^{p_n}(g_i^n)^2$. Similarly in an aforementioned way of kernel smoothing, an estimator of the spot volatility
can be obtained as
\begin{equation}
\widehat{\sigma^2}_{kp_nl_n/n}^{n,Pre}(p_n,l_n)=  \dfrac{n}{p_nl_n\varphi_n} \sum_{j=kl_n+1}^{ (k+1)l_n}(\overline{Z}_{jp_n}^n)^2,
\qquad \text{for} \quad k=0,\cdots, \lfloor n/(p_nl_n) \rfloor-1,
\end{equation}
where $l_n$ is the widow width of the kernel estimation.

In view of (\ref{model3}), we have $\overline{Z}_{jp_n}^n = \overline{X}_{jp_n}^n + \overline{\epsilon}_{jp_n}^n$. Some simple
variance calculations show that $\overline{X}_{jp_n}^n = O_p(\sqrt{\dfrac{p_n}{n}})$ and $\overline{\epsilon}_{jp_n}^n = O_p(\sqrt{\dfrac{1}{p_n}})$.
If $p_n\rightarrow \infty$ and $p_n^2/n \rightarrow \infty$ hold, as $n\rightarrow \infty$, then $\overline{Z}_{jp_n}^n$ is dominated
by $\overline{X}_{jp_n}^n$, and the effect of the market microstructure noise can then be neglected. As a consequence, we have $\widehat{IV}^{n,Pre}(p_n) \rightarrow^{p} \int_{0}^{1}\sigma_s^2ds$ and $\widehat{\sigma^2}_{kp_nl_n/n}^{n,Pre}(p_n,l_n) \rightarrow^{p} \sigma_{kp_nl_n/n}^2$.
If further that the conditions $p_n^5/n^3$ and $\sqrt{nl_n}/p_n \rightarrow 0$ are satisfied,  we have the following central limit
theorems:
\begin{align}\label{noi_clt1}
& \sqrt{\dfrac{n}{p_n}}(\widehat{IV}^{n,Pre}(p_n)- \int_{0}^{1}\sigma_s^2ds) \rightarrow^{ds} N(0,\int_{0}^{1}2\sigma_s^4ds),\\\label{noi_clt2}
&\sqrt{l_n}(\widehat{\sigma^2}_{kp_nl_n/n}^{n,Pre}(p_n,l_n) - \sigma_{kp_nl_n/n}^2) \rightarrow^{ds} N(0,2\sigma_{kp_nl_n}^4).
\end{align}
We also give a sketch of their proofs in Appendix (Lemma \ref{lem1}) as a by-product of this paper. Based on these results, we can establish our test procedure as 

\begin{thm}\label{thm3}
$X$ follows the process in \eqref{model1}, and Assumptions \ref{asu3}-\ref{asu4} hold.
\begin{enumerate}
\item For $ \omega \in \Omega$, if  as $n \rightarrow \infty$, $p_n \rightarrow \infty$, $l_n \rightarrow \infty$, and $p_n^{2}/n \rightarrow \infty$ hold, then we have, as $n \rightarrow \infty$, 
	\begin{equation}\label{thm3_1}
	\dfrac{p_nl_n}{n} \sum_{k=0}^{\lfloor n/(p_nl_n) \rfloor-1 } (\widehat{\sigma^2}_{kp_nl_n/n}^{n,Pre}(p_n,l_n)-\widehat{IV}^{n,Pre}(p_n) )^2 \rightarrow^{p} \int_{0}^{1} (\sigma_s^2 - IV)^2ds.
	\end{equation} 
\item For $\omega \in \Omega^{c}$, if as $n \rightarrow \infty$, $p_n \rightarrow \infty$, $l_n \rightarrow \infty$, $\sqrt{nl_n}/p_n\rightarrow 0$ and $p_n^{5}/n^3 \rightarrow 0$ hold, then we have, as $n \rightarrow \infty$, 
	\begin{equation}\label{thm3_2}
	\mathcal{T}^{n,Pre}(p_n,l_n):= \sqrt{\dfrac{p_nl_n}{2n}} \sum_{k=0}^{\lfloor n/(p_nl_n) \rfloor-1 } \Big( \big(\dfrac{\sqrt{l_n}(\widehat{\sigma^2}_{kp_nl_n/n}^{n,Pre}(p_n,l_n)-\widehat{IV}^{n,Pre}(p_n) )}{\sqrt{2}\widehat{IV}^{n,Pre}(p_n)}\big)^2 - 1\Big) \rightarrow^{d} N(0,1).
	\end{equation} 
\item Denote $ z_{\alpha} $ as the $\alpha$-quantile of standard normal distribution, if as $n \rightarrow \infty$, $p_n \rightarrow \infty$, $l_n \rightarrow \infty$, $\sqrt{nl_n}/p_n\rightarrow 0$ and $p_n^{3/2}/n \rightarrow 0$ hold, then we have, as $n \rightarrow \infty$, 
	\begin{align}
	\begin{cases}
	&\mathcal{P}( \mathcal{T}^{n,Pre}(p_n,l_n) > z_{1-\alpha}|\Omega^{c}) \rightarrow \alpha, \ \text{if} \ \mathcal{P}(\Omega^{c}) > 0, \\
	&\mathcal{P}( \mathcal{T}^{n,Pre}(p_n,l_n) > z_{1-\alpha}|\Omega^{v}) \rightarrow 1.
	\end{cases}
	\end{align}
	\end{enumerate}
\end{thm}

On one hand, we consider constructing our estimators of the integrated volatility and the spot volatility by using non-overlapping pre-averaged data for simplicity, instead of the overlapping case considered in \citet{JLMPV2009}. 
It has no harm to our theoretical results, but at a cost of reducing the number of pre-averaged data. 
On the other hand, as mentioned above, we diminish the effect of the noise by choosing $p_n^2/n \rightarrow \infty$. Alternatively, we can also take $p_n = O_p(n^{1/2})$, then $\overline{X}_{jp_n}^n$
and $\overline{\epsilon}_{jp_n}^n $ are of the same order. 
In this case, the effect of the noise should be removed by subtracting an estimator of the variance of the noise. Furthermore, the presence of the noise also deforms the variances of the asymptotic distributions
in (\ref{noi_clt1}) and (\ref{noi_clt2}), thus new estimators of these variances are necessarily to be reconstructed. 
It is viable to extend our test procedure to the setting with $p_n = O_p(n^{1/2})$ and overlapping pre-averaged data, but such a
consideration can complicate our test procedure to an undesirable degree. We mention that such a setting may be considered as a sole work for
our future research. Readers who are interested in this setting can refer to \citet{JLMPV2009} for the detailed discussion when it
comes to the estimation of the integrated volatility.

\begin{rmk}
	If the presence of jump process and  market microstructure noise are both considered simultaneously, we can obtain similar results by combining the thresholding technique and the pre-averaging method. The extension can be obviously seen from our previous derivation, thus we omit the detailed discussion here. Related papers can be
referred to are \citet{JLK2014} and references therein.
\end{rmk}

\section{Monte Carlo study}\label{sec3}
We now conduct some Monte Carlo simulation studies to examine our test procedure and investigate the finite sample performance of our test estimator in the cases of constant volatility and stochastic volatility. 
As discussed in the last theoretical section, we consider three different scenarios where continuous semi-martingale, involvement of finite jumps and contamination from market microstructure noise are used for modeling the log-price process. For the notations, we follow their definitions given in previous sections for the old ones, and shall specify later where new ones are used.

\subsection{Simulation design}
The latent log-price process $X=(X_{t})_{0 \leq t\leq 1}$ is generated from the following two stochastic differential equations, one of them considers constant volatility while the other one considers stochastic volatility.

$ \bullet $ Model 1--The constant volatility model
\begin{gather}
dX_{t}=\sigma dW_t,
\end{gather}
with $X_0=1$ and $\sigma=1$.

$ \bullet $ Model 2--The Heston model with stochastic volatility
\begin{align}
\begin{split}
dX_{t}&=\sigma_t dW_t,\\
d\sigma_{t}^{2}&=\kappa(\alpha-\sigma_{t}^{2})dt+\gamma\sigma_{t}(\rho dW_t+\sqrt{1-\rho^2}dB_t),
\end{split}
\end{align}
with the parameters $\kappa=5, \alpha=0.04,\gamma=5$, $\rho=-\sqrt{0.5}$, $X_0=1$ and $\sigma_0=1$. We follow the parameter setting in
\cite{WM2014} to calibrate the model to real financial data.

Regarding the jump component $J_t = \sum_{j=1}^{N_t} \gamma_{\tau_j}$ in Section \ref{part:jump}, we consider the jump size $\gamma_{\tau_j} \sim N(0,\sigma_{\kappa}^{2})$, the number of jumps up to time point $t$, $N_{t}\sim Poisson(\lambda t)$, which is a Poisson distribution with
parameter $\lambda$. We firstly generate the process $N_{t}$ within $t\in[0,1]$, and in subsequence generate $\gamma_{\tau_j}$ independently.
We fix $\sigma_{\kappa}=0.5$ and choose different jump intensity by setting the parameter $\lambda=10, 20, 50, 100$. For estimating the spot
volatility, the window-width is set as $k_n=\lfloor\frac{\theta}{\sqrt{\Delta_n}}\rfloor$ with $\theta=1.2$ for satisfying the theoretical
conditions in Theorem \ref{thm2}. We apply the thresholding technique to filter the jumps by setting the truncation level $\nu_n=4\sqrt{BV_n} \cdot \Delta_n^{\varpi}$ with $\varpi=0.499$ and
\begin{align*}
BV_n=\frac{\pi}{2}\sum\limits_{i=2}^{n} |\Delta_i^nY||\Delta_{i-1}^nY|.
\end{align*}
The quantity $BV_n$ is the realized bipower variation estimator introduced in \citet{BN2004} and serves as a consistent estimator of the integrated
volatility which is robust to jumps.

For the market microstructure noise term $\epsilon_t$ in Section \ref{part:noise}, it is mixed in the observed prices at $t= 0, 1/n, ..., n/n$, with $\epsilon_{t} \sim N(0,\eta^2)$. The noise terms are independently and identically distributed with different strengths $\eta=0.001, 0.01, 0.05$. Recall that $p_n$ is the number of increments based on the raw data used for pre-averaging, $l_n$ is the number of non-overlapping pre-averaged blocks used for the kernel estimation of the spot volatility. For Theorem \ref{thm3}, we take $p_n = \lfloor c \Delta_n^{-(1/2+\chi)}\rfloor$ with $c=1/3$ and $\chi=0.05$, and $l_n = \lfloor a \Delta_n^{-b} \rfloor$ with $a = 2$ 
and $b = 0.17$, which satisfy our theoretical requirement.

For each experiment, we simulate 5000 runs of daily sample paths by using the Euler discretization method. We consider different sampling 
frequencies with $ n= 23400, 11700, 7800$, $4680, 2340, 1170, 780$, corresponding to sampling at every 1, 2, 3, 5, 10, 20, 30 seconds respectively, 
over a 6.5-hour trading day in the U.S. stock market.

\subsection{Simulation results}
Table \ref{tab1} records the empirical size (based on constant volatility) and power (based on time-varying volatility) of the heteroscedasticity
test when finite activity jumps are present in the logarithmic price process. The setting $\lambda = 0$, which corresponds to the continuous semi-martingale model without jumps, is also documented for comparison. We observe desirable size performances, meaning that the probability 
of type I error is acceptable, for $\lambda=0, 10, 20$ and all $n$ considered. For fixed $\lambda$, the magnitude of size approaches to corresponding
nominal confidence level as the sampling frequency increases, this is even more evident for relatively larger $\lambda$. As for the influence of jumps, we see that more intensive jumps always worsen the performance of size, and the extent is more obvious when the sample size $n$ is relatively 
smaller. We find that almost all the values of power are 1 for all $n$, $\lambda$ and the three nominal levels, which shows our test is quite 
powerful in detecting the time variation in volatility process. This is inline with our theoretical analysis in Section \ref{part:cont} that 
our test estimator diverges at a fast rate if the volatility process is not constant.

\begin{table}[!htbp]
\centering
		\begin{tabular}{lllllllllllllllllll}\toprule			
				\multirow{2}{*}{} & \multirow{2}*{n}  & \multicolumn{3}{c}{Size} & \multicolumn{3}{c}{Power}\\
				\cmidrule(r){3-5} \cmidrule(r){6-8} \cmidrule(r){9-11}
				& &10\% & 5\%& 1\% &10\% & 5\%& 1\% \\
				\hline																				
				&	780 	&0.0882	&0.0382 &0.0096 &1.0000	&1.0000	&1.0000	\\
				\multirow{2}{*}{$\lambda=0$}
				&	2340 	&0.0920	&0.0422 &0.0096 &1.0000	&1.0000	&1.0000 \\
				&	7800 	&0.1038	&0.0536 &0.0118 &1.0000	&1.0000	&1.0000	\\
				&	23400 	&0.1006	&0.0460 &0.0100 &1.0000	&1.0000	&1.0000	\\
				\hline																			
				&	780 	&0.1050	&0.0586 &0.0188 &1.0000	&1.0000	&1.0000 \\
				\multirow{2}{*}{$\lambda=10$}
&	2340 	&0.0918	    &0.0438      &0.0104      &1.0000   	&1.0000 	    &1.0000\\
&	7800 	&0.0988	    &0.0538      &0.0130      &1.0000   	&1.0000 	    &1.0000\\
&	23400 	&0.0992	    &0.0512      &0.0090      &1.0000   	&1.0000 	    &1.0000\\
\hline	
&	780 	&0.1270	    &0.0794      &0.0322      &1.0000   	&1.0000 	    &1.0000\\
\multirow{2}{*}{$\lambda=20$}
&	2340 	&0.1024	    &0.0564      &0.0156      &1.0000   	&1.0000 	    &1.0000\\
&	7800 	&0.1004	    &0.0526      &0.0142      &1.0000   	&1.0000 	    &1.0000\\
&	23400 	&0.1008	    &0.0488      &0.0094      &1.0000   	&1.0000 	    &1.0000\\
\hline	
&	780 	&0.3422	    &0.2710      &0.1734      &1.0000   	&1.0000 	    &0.9998\\
\multirow{2}{*}{$\lambda=50$}
&	2340 	&0.1496	    &0.0934      &0.0420      &1.0000   	&1.0000 	    &1.0000\\
&	7800 	&0.1084	    &0.0556      &0.0140      &1.0000   	&1.0000 	    &1.0000\\
&	23400 	&0.0900	    &0.0464      &0.0078      &1.0000   	&1.0000 	    &1.0000\\
\hline	
&	780 	&0.8048	    &0.7572      &0.6478      &1.0000   	&1.0000 	    &1.0000\\
\multirow{2}{*}{$\lambda=100$}
&	2340 	&0.4574	    &0.3806      &0.2520      &1.0000   	&1.0000 	    &1.0000\\
&	7800 	&0.1520	    &0.0968      &0.0408      &1.0000   	&1.0000 	    &1.0000\\
&	23400 	&0.0984	    &0.0514      &0.0132      &1.0000   	&1.0000 	    &1.0000\\
				\bottomrule
		\end{tabular}\\
     \caption{Heteroscedasticity test with jumps. } \label{tab1}
\end{table}

Table \ref{tab2} documents the size and power of the heteroscedasticity test in the presence of market microstructure noise. The finite sample performances of both size and power are satisfying for $\eta = 0.001, 0.01$. For fixed $\eta$, as the sampling  frequency increases, the size 
and power perform better in the sense of getting closer to corresponding nominal confidence levels and 1, respectively. This phenomenon is even 
more distinct for relatively larger $\eta$. Regarding the effect of the market microstructure noise, a larger $\eta$ deviates the values of Power away from 1 and yields a larger type II error. Moreover, such a deterioration is even worse for relatively smaller sample size $n$. We also find that 
all the values of size are close to corresponding nominal confidence levels for all the different parameters $n$ and $\eta$ considered. This 
justifies that the pre-averaging technique works well in demolishing the disturbance from the market microstructure noise for the estimation of 
volatility (the integrated volatility or/and the spot volatility). The results of power are more sensitive to the presence of market microstructure 
noise because our statistics $\mathcal{T}^{n,Pre}(p_n,l_n)$ in Theorem \ref{thm3} diverges to infinity in a relatively slow rate which depends on the parameters $p_n,l_n$, when the constant volatility assumption is violated.

\begin{table}[ht]
	\begin{center}{
			\begin{tabular}{lllllllllllllllllll}\toprule
				
				\multirow{2}{*}{} & \multirow{2}*{n}  & \multicolumn{3}{c}{Size} & \multicolumn{3}{c}{Power}\\
\cmidrule(r){3-5} \cmidrule(r){6-8} \cmidrule(r){9-11}
 & &10\% & 5\%& 1\% &10\% & 5\%& 1\% \\
		\hline														
\multirow{4}{*}{$\eta=0.001$}	
&	1170 	&0.0956  	&0.0510      &0.0224	            &0.9976      &0.9968	    &0.9936\\				
&	4680 	&0.1078  	&0.0576      &0.0222	            &0.9994      &0.9992	    &0.9988\\
&	11700 	&0.1130  	&0.0598      &0.0210	            &0.9996      &0.9996        &0.9992\\
&	23400 	&0.1034  	&0.0528      &0.0166	            &1.0000      &1.0000	    &0.9996 \\
\hline	
\multirow{4}{*}{$\eta=0.01$}
&	1170 	&0.0932  	&0.0504      &0.0204	            &0.9934      &0.9912	    &0.9852\\														
&	4680 	&0.1008  	&0.0554      &0.0194	            &0.9994      &0.9992        &0.9992 \\
&	11700 	&0.1108  	&0.0570      &0.0170	            &0.9996      &0.9994	    &0.9994 \\
&	23400 	&0.1094  	&0.0548      &0.0162	            &0.9996      &0.9996	    &0.9996 \\
\hline	
\multirow{4}{*}{$\eta=0.05$}
&	1170 	&0.0922  	&0.0450      &0.0216	            &0.6918      &0.6586	    &0.6052\\														
&	4680 	&0.1114  	&0.0576      &0.0192	            &0.8286      &0.8072 	    &0.7652\\
&	11700 	&0.1104  	&0.0568      &0.0182	            &0.9074      &0.8928	    &0.8698 \\
&	23400 	&0.1064  	&0.0562      &0.0146	            &0.9366      &0.9282	    &0.9060 \\
\bottomrule
\end{tabular}}
\end{center}
\caption{Heteroscedasticity test with market microstructure noise. }\label{tab2}
\end{table}

\begin{figure}[!htbp]
	\centering
	\subfigure[Histogram]{
		\begin{minipage}[b]{0.45\textwidth}
			\includegraphics[width=1.2\textwidth]{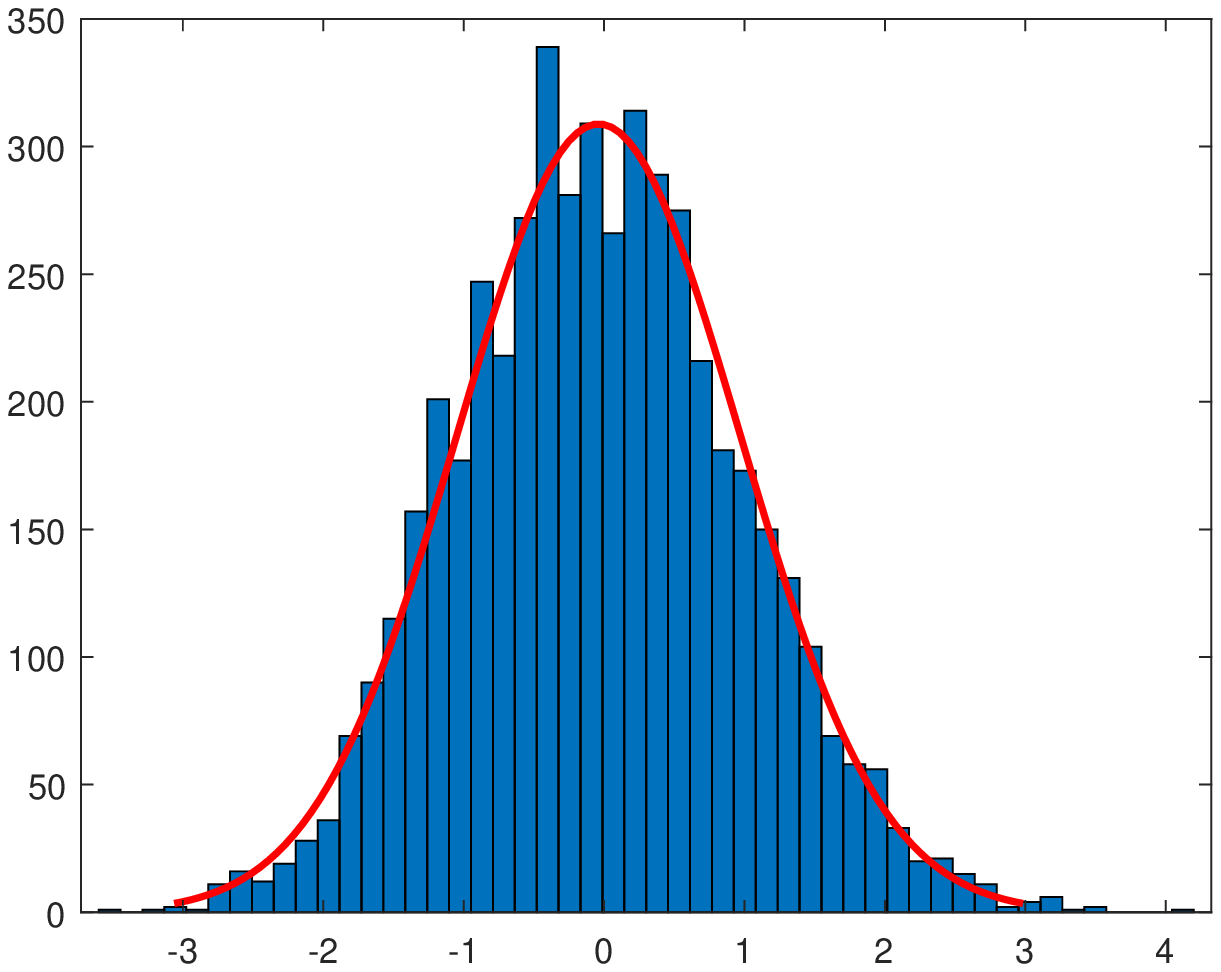}
		\end{minipage}
	}
	\subfigure[Q-Q Plot]{
		\begin{minipage}[b]{0.45\textwidth}
			\includegraphics[width=1.2\textwidth]{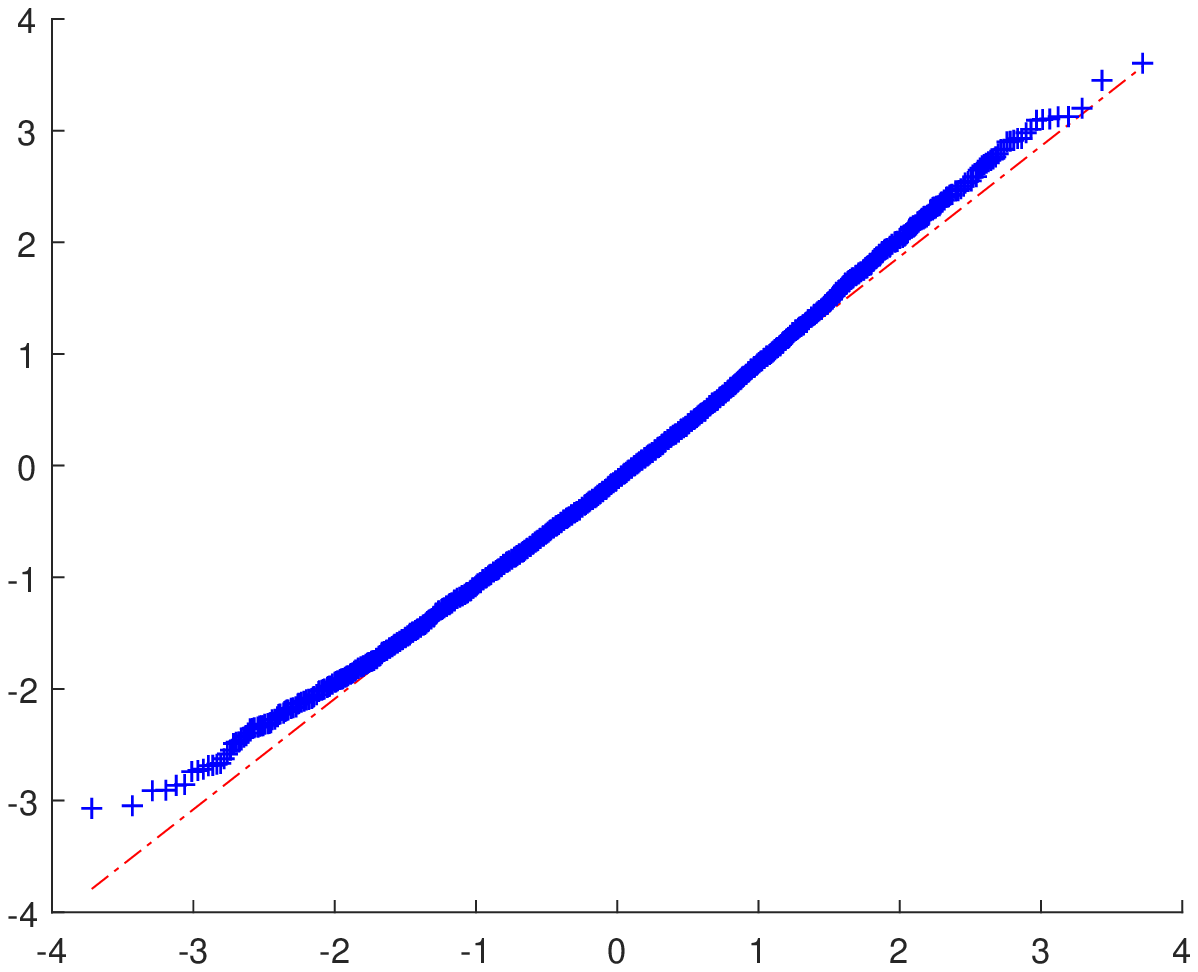}
		\end{minipage}
	}
	\caption{Estimates of $\mathcal{T}^{n}(k_n)$ in \eqref{thm1_2} of Theorem \ref{thm1} with $n = 23400$. In the histograms, the red real curve is the density of standard normal random variable.
	}\label{fig:QQPlot1}
\end{figure}

\begin{figure}[!htbp]
	\centering
	\subfigure[Histogram]{
		\begin{minipage}[b]{0.45\textwidth}
			\includegraphics[width=1.2\textwidth]{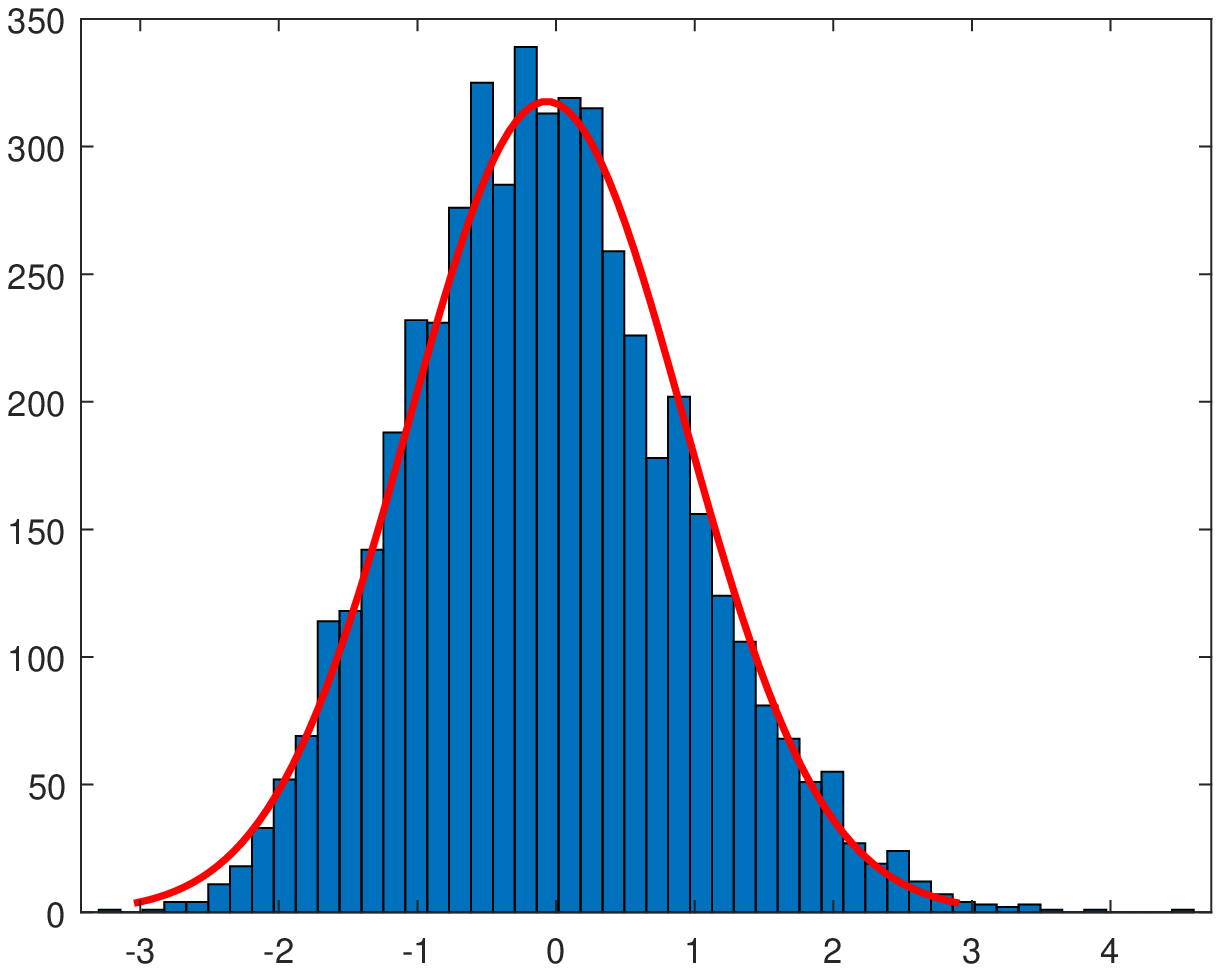}
		\end{minipage}
	}
	\subfigure[Q-Q Plot]{
		\begin{minipage}[b]{0.45\textwidth}
			\includegraphics[width=1.2\textwidth]{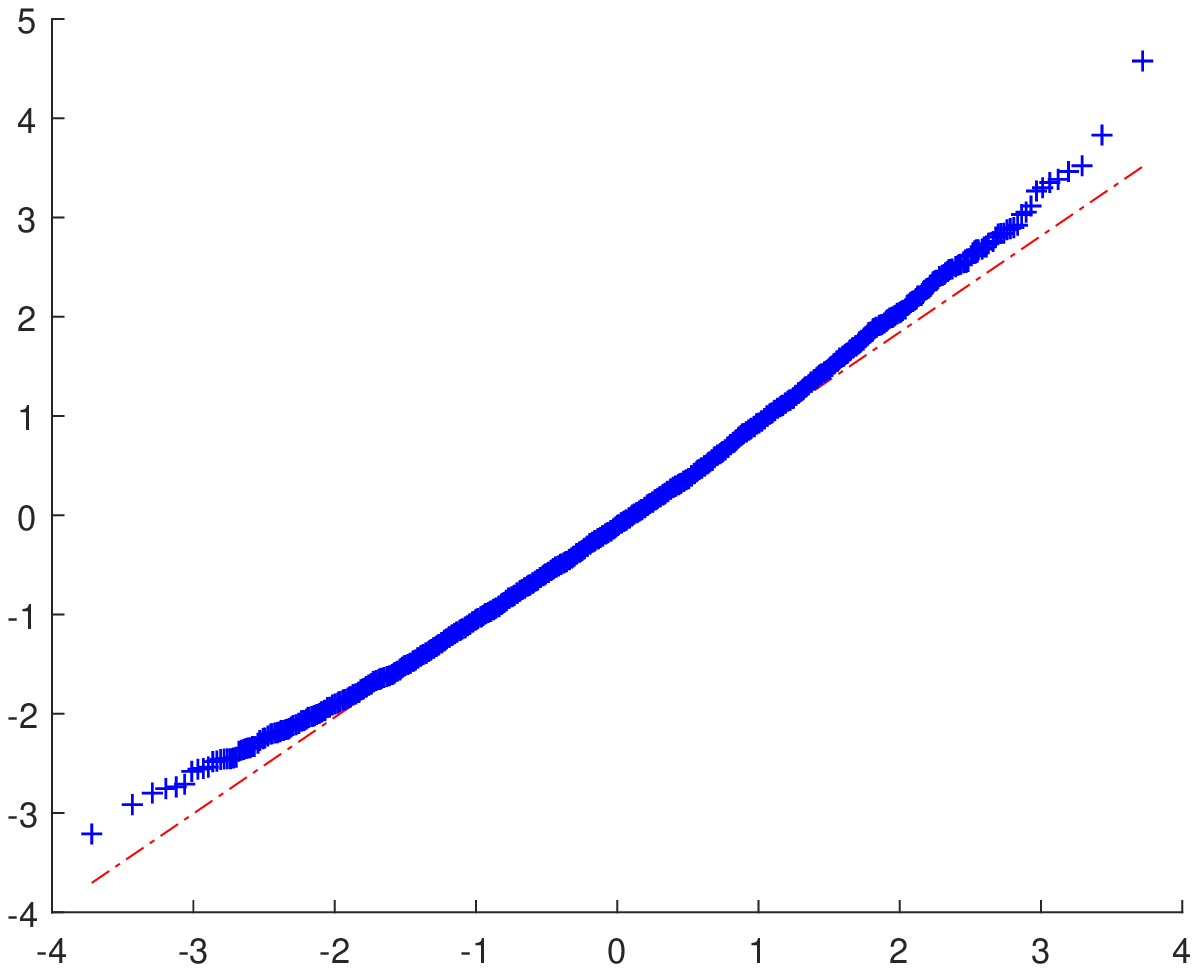}
		\end{minipage}
	}
	\caption{Estimates of $\mathcal{T}^{n,Thr}(k_n)$ in \eqref{thm2_2} of Theorem \ref{thm2}, with $\lambda = 20$ and $n = 23400$. In the histograms, the red real curve is the density of standard normal random variable.
	}\label{fig:QQPlot2}
\end{figure}

\begin{figure}[!htbp]
	\centering
	\subfigure[Histogram]{
		\begin{minipage}[b]{0.45\textwidth}
			\includegraphics[width=1.2\textwidth]{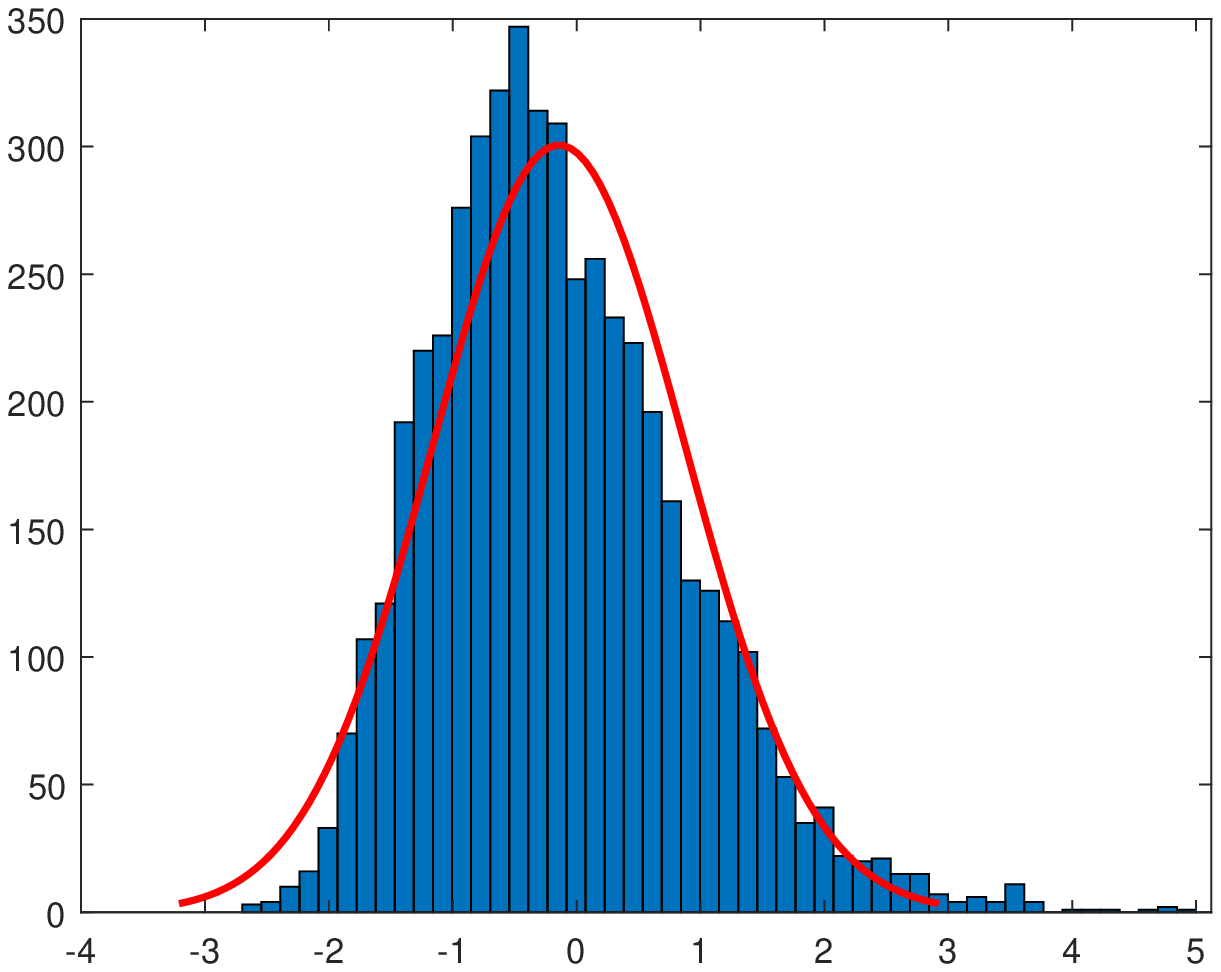}
		\end{minipage}
	}
	\subfigure[Q-Q Plot]{
		\begin{minipage}[b]{0.45\textwidth}
			\includegraphics[width=1.2\textwidth]{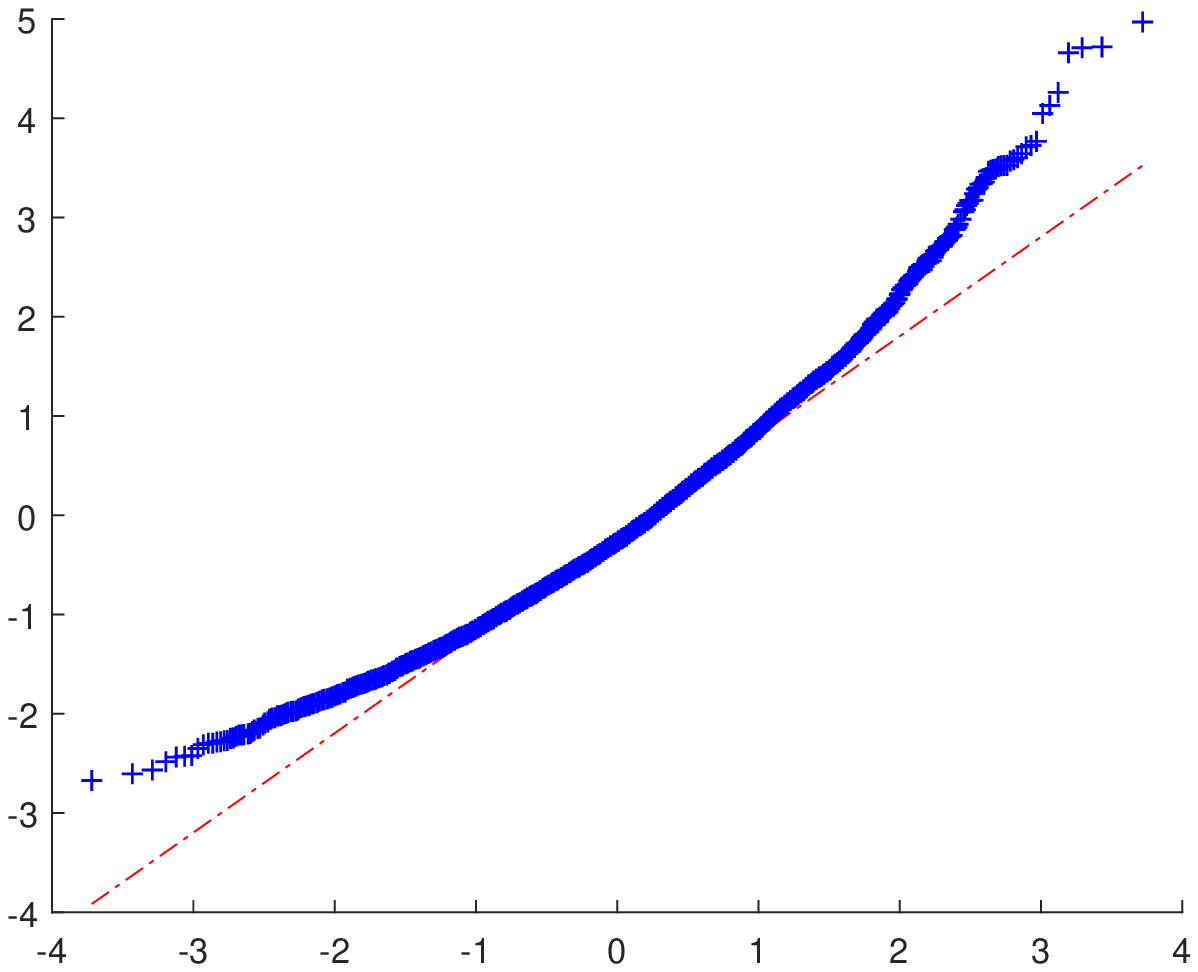}
		\end{minipage}
	}
	\caption{Estimates of $\mathcal{T}^{n,Pre}(p_n,l_n)$ in \eqref{thm3_2} of Theorem \ref{thm3}, with $\eta = 0.01$ and $n = 23400$. In the histograms, the red real curve is the density of standard normal random variable.
	}\label{fig:QQPlot3}
\end{figure}

To verify the accuracy of the normal approximations of our test statistics, namely \eqref{thm1_2} in Theorem \ref{thm1}, \eqref{thm2_2} in Theorem \ref{thm2} and \eqref{thm3_2} in Theorem \ref{thm3}, 
we demonstrate Q-Q plots and histograms for the finite estimates of $\mathcal{T}^{n}(k_n)$, $\mathcal{T}^{n,Thr}(k_n)$ and $\mathcal{T}^{n,Pre}(p_n,l_n)$ under the constant volatility model in Figures \ref{fig:QQPlot1}--\ref{fig:QQPlot3}. It is
shown that all the histograms approximate standard normal distribution closely and the Q-Q plots are almost linear, which proves the asymptotic normality of these three quantities.

\section{Real data analysis}\label{sec4}
In this section, we apply our proposed heteroscedasticity test statistics to high-frequency data from the NYSE TAQ database.
We use the transaction price data of the International Business Machines (IBM) in the whole year of 2011, with a total of 252
trading days. 
For various reasons, raw trading data contains numerous errors. Therefore, the data is not immediately suitable for analysis and data-cleaning is an essential step when dealing with tick-by-tick data. Following the pre-filtering routine of \citet{BNHLS2009}, we collect all transactions from 9:30 to 16:00, delete entries with zero prices, merge multiple transactions with the same time stamp by taking the weighted average of all prices and sample every 5 seconds in calendar time.
We consider 5-minute data for $\mathcal{T}^{n}(k_n)$ in Theorem \ref{thm1} and $\mathcal{T}^{n,Thr}(k_n)$ in Theorem \ref{thm2}
to avoid the influence of market microstructure noise, and 5-second data for $\mathcal{T}^{n,Pre}(p_n,l_n)$ in Theorem \ref{thm3}.

\begin{figure}[!htbp]
	\centering
	\subfigure[Without removing jumps and microstructure noise: based on $\mathcal{T}^{n}(k_n)$ in Theorem \ref{thm1} by using 5-minute data. ]{
		\begin{minipage}[b]{0.3\textwidth}
			\includegraphics[width=1.1\textwidth]{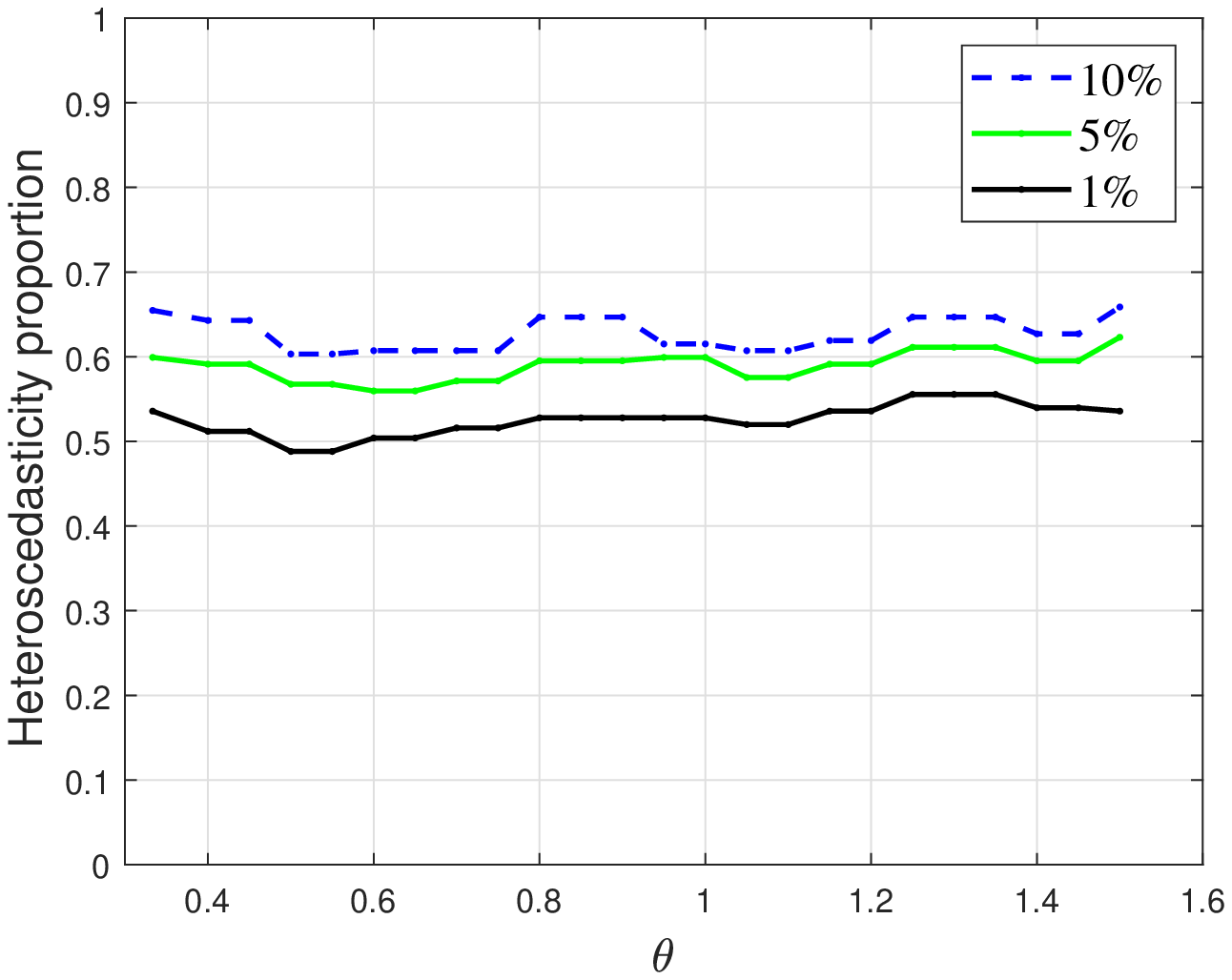}
		\end{minipage}
	}
	\subfigure[Removing jumps: based on $\mathcal{T}^{n,Thr}(k_n)$ in Theorem \ref{thm2} by using 5-minute data. ]{
		\begin{minipage}[b]{0.3\textwidth}
			\includegraphics[width=1.1\textwidth]{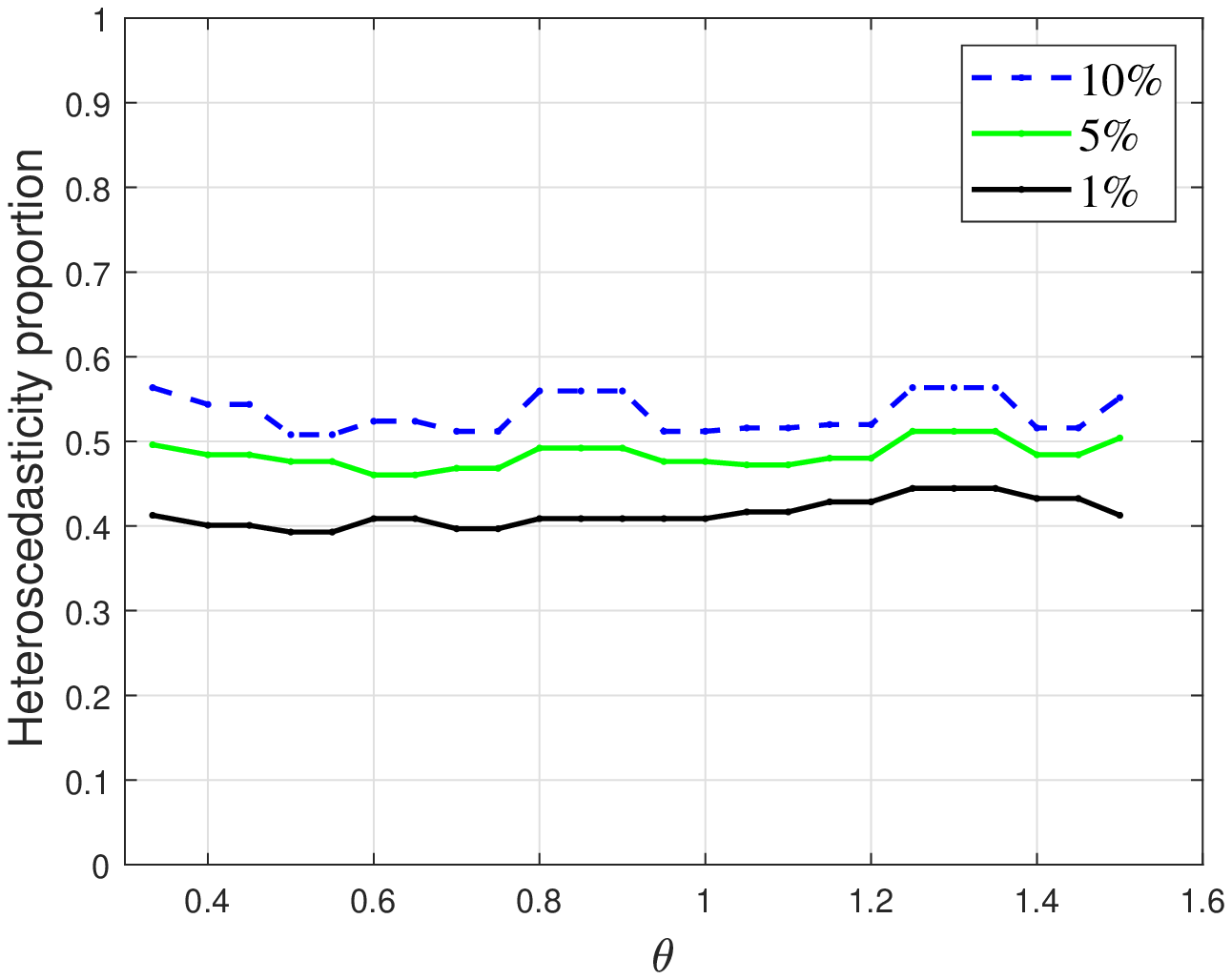}
		\end{minipage}
	}
	\subfigure[Removing microstructure noise: based on  $\mathcal{T}^{n,Pre}(p_n,l_n)$ in Theorem \ref{thm3} by using 5-second data. ]{
		\begin{minipage}[b]{0.3\textwidth}
			\includegraphics[width=1.1\textwidth]{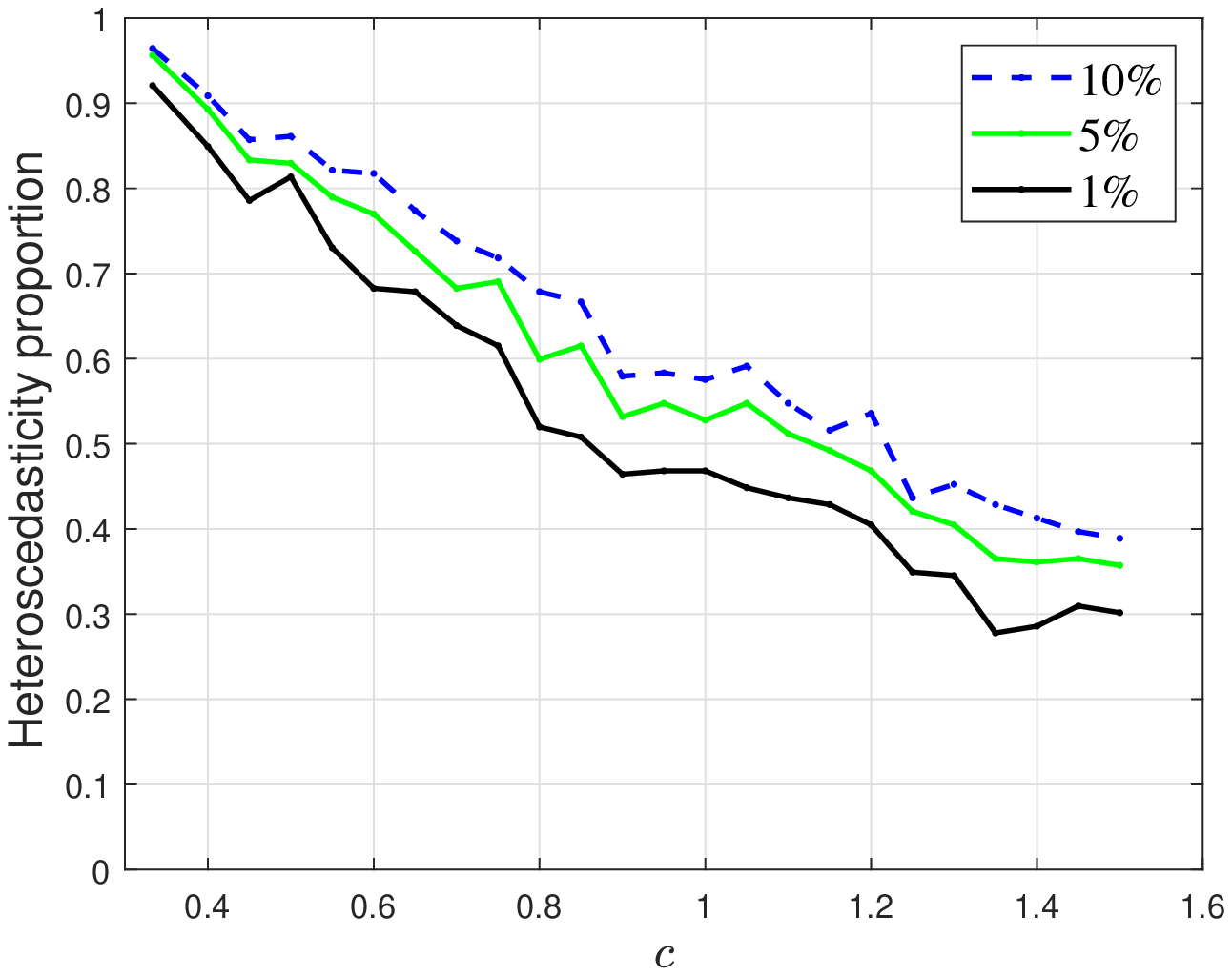}
		\end{minipage}
	}
	\caption{Test for the constancy of daily volatility for IBM in 2011.}\label{fig:Heteroscedasticityproportion}
\end{figure}

Recall that we set $k_n=\lfloor\frac{\theta}{\sqrt{\Delta_n}}\rfloor$ and $p_n = \lfloor c \Delta_n^{-(1/2+\chi)}\rfloor$ for the number of raw data used for kernel smoothing in the noise-free setting and number of raw data used for pre-averaging in the noisy setting, respectively.
Figure \ref{fig:Heteroscedasticityproportion} depicts the proportion of the day with time-varying volatility tested at significance levels of 10\%, 5\% and 1\% as a function of $\theta$ in the frictionless cases and a function of $c$ in the noisy setting. 
Regarding other related parameters not mentioned, they remain the same as the ones in the simulation section.
When we implement the test by using 5-minute high-frequency sampling to diminish the influence from the market microstructure noise, the proportions of heteroscedasticity volatility are insensitive to the choice of $\theta$. 
For the three different significance levels, similar patterns are observed for each scenario with small deviation in the magnitude of heteroscedasticity proportion for all range of $\theta$. 
If we remove the jumps by the truncation method, the proportions reduce by around 10\% for all the three significance levels compared to the case without removing the jumps. 
This is inline with the intuition that the presence of jumps makes the price process more volatile.
For the scenario of considering removing the market microstructure noise by using pre-averaged 5-second data, the heteroscedasticity proportion 
decreases as $c$ increases. 
This verifies that the pre-averaging methodology mitigates the impact of the market microstructure noise better for relative larger $c$, which corresponds to the case that more raw data are used for pre-averaging. 

\begin{figure}[!htbp]
	\centering
	\subfigure[Without removing jumps and microstructure noise: $\widehat{\sigma^2}_{\tau}^{n}(k_n)$ in Theorem \ref{thm1} by using 5-minute data. ]{
		\begin{minipage}[b]{0.3\textwidth}
			\includegraphics[width=1.1\textwidth]{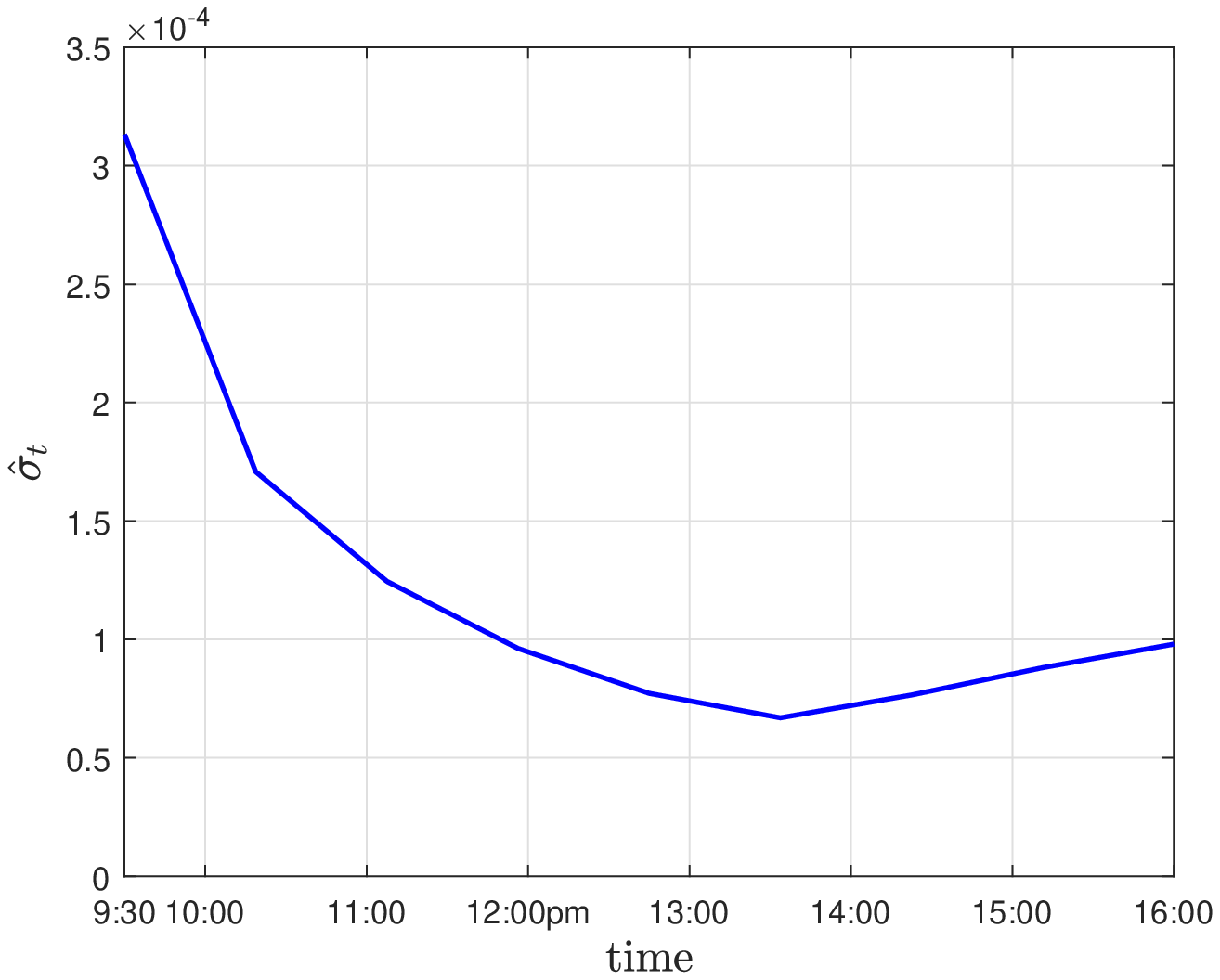}
		\end{minipage}
	}
	\subfigure[Removing jumps: $\widehat{\sigma^2_\tau}^{n,Thr}(k_n)$  in Theorem \ref{thm2} by using 5-minute data. ]{
		\begin{minipage}[b]{0.3\textwidth}
			\includegraphics[width=1.1\textwidth]{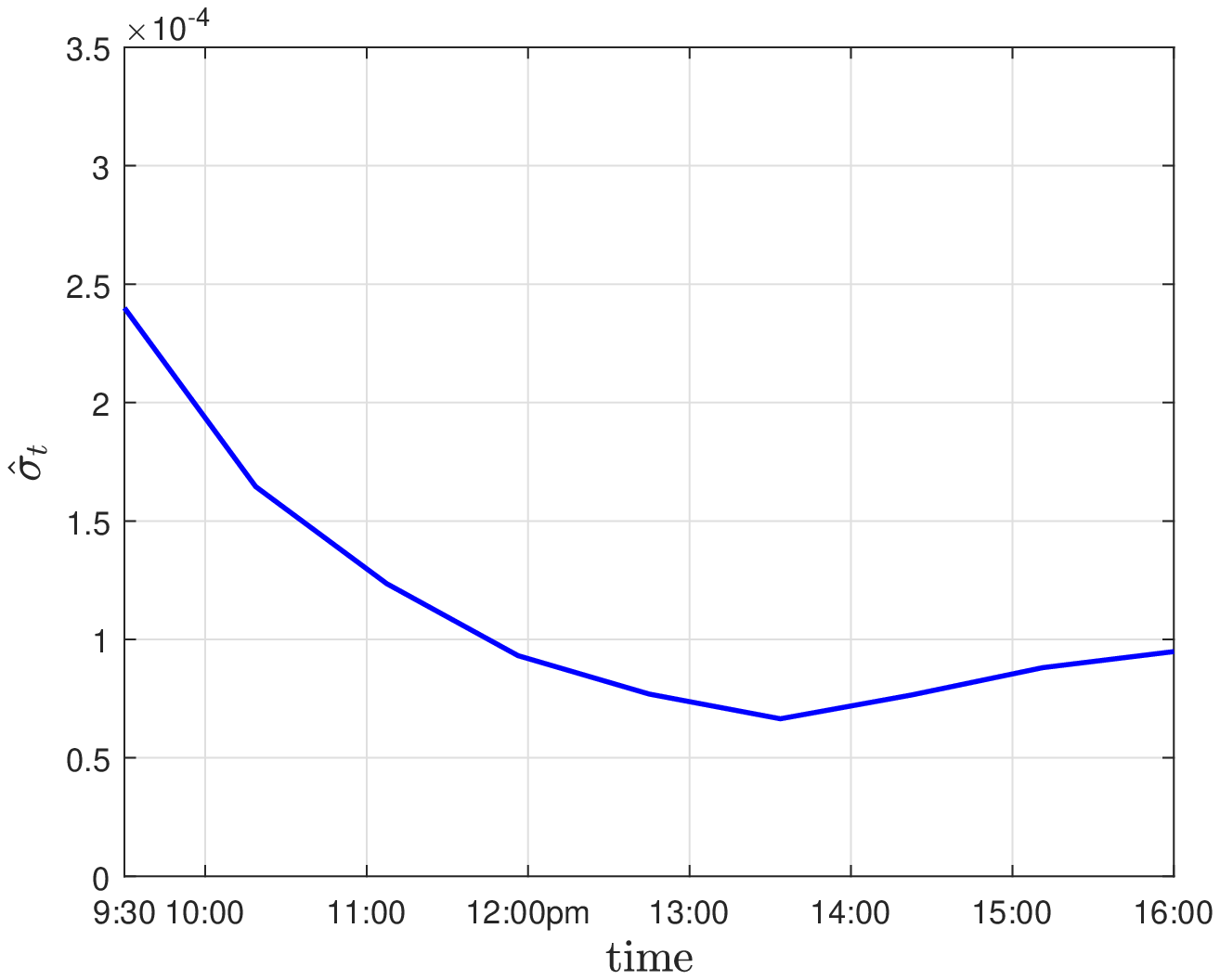}
		\end{minipage}
	}
	\subfigure[Removing microstructure noise: $\widehat{\sigma^2}_{kp_nl_n/n}^{n,Pre}(p_n,l_n)$ in Theorem \ref{thm3} by using 5-second data. ]{
		\begin{minipage}[b]{0.3\textwidth}
			\includegraphics[width=1.1\textwidth]{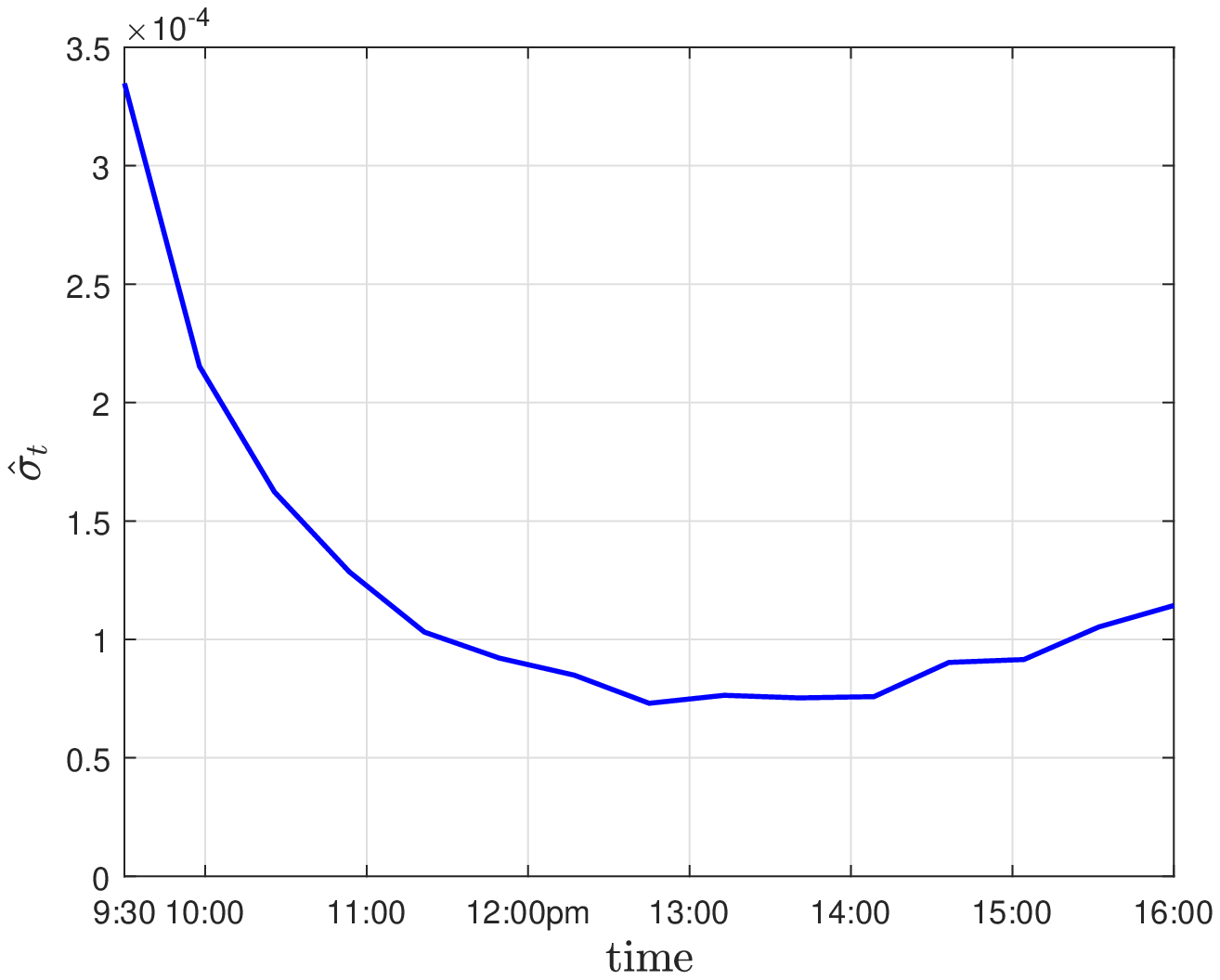}
		\end{minipage}
	}
	\caption{The cross-sectional average of intraday volatility curves estimated by the spot volatility estimators for IBM. }\label{fig:volcur}
\end{figure}

In Figure \ref{fig:volcur}, we demonstrate the cross-sectional average of intraday volatility curves estimated by the spot volatility estimators $\widehat{\sigma^2}_{\tau}^{n}(k_n)$ in Theorem \ref{thm1}  for the continuous setting, $\widehat{\sigma^2_\tau}^{n,Thr}(k_n)$  in Theorem \ref{thm2} for the setting with jumps, and $\widehat{\sigma^2}_{kp_nl_n/n}^{n,Pre}(p_n,l_n)$ in Theorem \ref{thm3} for the noisy setting respectively.
It is shown that the volatility estimates near the opening time or the closing time are relatively larger than other time points in the middle time span. Moreover, the estimated volatility curves are roughly with sharp decreases or increases around pre-scheduled macroeconomic announcements (e.g., at 10:00 or 14:00). 
This makes the whole volatility curve like 
a reverted ``J"-shape, which is also found in \cite{christensen2018diurnal}. 
In fact, this happens for most of the days in a year. 
There are also empirical literatures explaining the phenomenon. 
For example, the period covering 9:30 and 10:00 is associated with market-wide news such as FOMC meetings and macroeconomic reports, which make the stock prices to be more volatile, as discussed in \citet{LM2008}, \citet{Lee2012}, and etc.
\begin{table}[!htbp]
	\centering
	\begin{tabular}{llllllllllllllllllll}\toprule			
		\multirow{2}{*}{} & \multicolumn{4}{c}{$\mathcal{T}^{n}(k_n)$} & \multicolumn{3}{c}{$\mathcal{T}^{n,Thr}(k_n)$} & \multicolumn{3}{c}{$\mathcal{T}^{n,Pre}(p_n,l_n)$} \\
		\cmidrule(r){3-5} \cmidrule(r){6-8} \cmidrule(r){9-11}
		& &10\% & 5\%& 1\% &10\% & 5\%& 1\%  &10\% & 5\%& 1\% \\
		\hline																				
		 09:30-16:00 &	 &0.6151 &0.5992 &0.5278 &0.5119 &0.4762 &0.4087 &0.8214 &0.7897 &0.7302\\
		\hline																				
		10:00-15:30  &	 &0.4325 &0.3730 &0.3016 &0.4087 &0.3492 &0.2778 &0.7024 &0.6429 &0.5278\\
		\hline																				
		10:30-15:00  &	 &0.2778 &0.2302 &0.1587 &0.2778 &0.2302 &0.1587 &0.5913 &0.5198 &0.4643 \\
		\bottomrule
	\end{tabular}\\
	\caption{Heteroscedasticity test for different time spans, with $\theta = 1.0$ and $c = 1/3$. } \label{tab3}
\end{table}

To quantify how does the variation of stock price during the opening and closing time affect our heteroscedasticity test procedure, we record the heteroscedasticity proportion results within three different time spans, namely 09:30-16:00, 10:00-15:30 and 10:30-15:00, in Table \ref{tab3}. 
For all three significance levels of 10\%, 5\% and 1\%, we see that the proportions decrease as we gradually discard data obtained in the opening and closing periods for our heteroscedasticity test. 
The observation implies that the variation during the opening and closing periods leads to a test result of time-varying intraday volatility for most of the days tested.


\section{Conclusion}\label{sec5}
In this paper, we propose a new nonparametric way to do the heteroscedasticity test for high-frequency data. The test procedure is based on the estimations of integrated volatility and spot volatility, for which a great deal of existing
literatures can be found. Our test procedure is easy to conduct and can be naturally extended to different settings, such
as the cases in the presence of jumps and market microstructure noise. Our Monte Carlo simulation studies show the good finite sample performance of the asymptotic theory. Finally, we also apply our test procedure to do the heteroscedasticity test for some real high-frequency financial data. The empirical studies indicate that the volatility is not constant in most of days, and the opening and closing periods account for a relatively large proportion of intraday heteroscedasticity.
This paper also enlighten us on testing whether the covariance structure between different assets is constant or not during a given time interval, as a future work.

\section*{Acknowledgement}
Qiang Liu's work is supported by MOE-AcRF Grant of Singapore (No. R-146-000-258-114), Zhi Liu gratefully acknowledges financial support from FDCT of Macau (No. 202/2017/A3) and NSFC (No. 11971507), 
Chuanhai Zhang's research is supported in part by Humanity and Social Science Youth Foundation of Chinese Ministry of Education (No. 18YJC790210) and in part by the Fundamental Research Funds for the Central Universities, Zhongnan University of Economics and Law (2722019PY038).

\section*{Appendix}
For the following proofs, by a standard localization procedure given in \citet{BGJPS2006}, we can
replace the local boundedness hypothesis in our setting by a boundedness one without
loss of generality. We use an unified $C$ to denote positive constants in the proofs,
and it may change from line to line. Note that
\begin{equation*}
\Delta_i^n X = \int_{(i-1)/n}^{i/n} b_sds+ \int_{(i-1)/n}^{i/n} \sigma_sdB_s.
\end{equation*}
It is obvious that the random variable $\int_{(i-1)/n}^{i/n} b_sds$ is dominated by $\int_{(i-1)/n}^{i/n}
\sigma_sdB_s$, so the drift term $b_s$ has no effect on asymptotic properties of estimators where
$X$ is involved. Thus, setting $b_s\equiv 0$ has no influence on our theoretical results, and it
can simplify the whole proof procedure to some extends.

\textbf{Proof of Theorem \ref{thm1}: }
\textbf{(1)} Note that
\begin{align*}
& \dfrac{k_n}{n} \sum_{j=0}^{\lfloor n/k_n \rfloor-1 } (\widehat{\sigma^2}_{jk_n/n}^{n}(k_n)-\widehat{IV}^{n} )^2 - \int_{0}^{1} (\sigma_s^2 - IV)^2ds \\
& = \dfrac{k_n}{n} \sum_{j=0}^{\lfloor n/k_n \rfloor-1 } \big( (\widehat{\sigma^2}_{jk_n/n}^{n}(k_n)-\widehat{IV}^{n} )^2 - (\sigma^2_{jk_n/n}- IV )^2 \big) \\
& + \dfrac{k_n}{n} \sum_{j=0}^{\lfloor n/k_n \rfloor-1} (\sigma^2_{jk_n/n}- IV )^2 - \int_{0}^{1} (\sigma_s^2 - IV)^2ds \\
& := A_1 + A_2.
\end{align*}
The result $A_1 \rightarrow^{p} 0$ directly follows from $\widehat{\sigma^2}_{jk_n/n}^{n}(k_n) \rightarrow^{p} \sigma^2_{jk_n/n} $
and $ \widehat{IV}^{n} \rightarrow^{p} IV$, whose proofs are given below. Observing
\begin{align*}
& \widehat{\sigma^2}_{jk_n/n}^{n}(k_n) - \sigma^2_{jk_n/n} \\
&= \dfrac{1}{k_n\Delta_n}\sum_{i=jk_n + 1}^{(j+1)k_n } \big( (\Delta_i^n X)^2  - (\sigma_{jk_n/n}\Delta_i^n B)^2 \big) + \dfrac{1}{k_n\Delta_n}\sum_{i=jk_n + 1}^{(j+1)k_n } \big( (\sigma_{jk_n/n}\Delta_i^n B)^2 - \sigma^2_{jk_n/n}\Delta_n \big) \\
&:= B_1 + B_2.
\end{align*}
Since
\begin{align*}
\mathbf{E}[| (\Delta_i^n X)^2  - (\sigma_{jk_n/n}\Delta_i^n B)^2 |] &\leq C\mathbf{E}[|\Delta_i^n X - \sigma_{jk_n/n}\Delta_i^n B |] \\
& \leq C(\mathbf{E}[|\Delta_i^n X - \sigma_{jk_n/n}\Delta_i^n B |^2])^{1/2} \\
& \leq C\Delta_n^{3/2}\sqrt{k_n},
\end{align*}
together with Holder's inequality and It$\hat{\text{o}}$'s isometry, we obtain that $B_1 \rightarrow^{p} 0$.
Note that $\{ \big( (\sigma_{jk_n/n}\Delta_i^n B)^2 - \sigma^2_{jk_n/n}\Delta_n \big), \mathcal{F}_{jk_n/n} \}$ is a martingale
difference array, thus
\begin{align*}
\mathbf{E} [(B_2)^2] = \dfrac{1}{(k_n\Delta_n)^2}\sum_{i=jk_n + 1}^{(j+1)k_n } \mathbf{E} \big[ \big( (\sigma_{jk_n/n}\Delta_i^n B)^2 - \sigma^2_{jk_n/n}\Delta_n \big)^2 \big| \mathcal{F}_{jk_n/n} \big] = \frac{2\sigma^4_{jk_n/n}}{k_n}.
\end{align*}
By Chebyshev's inequality, we obtain $B_2 \rightarrow^{p} 0$, hence $\widehat{\sigma^2}_{jk_n/n}^{n}(k_n) \rightarrow^{p} \sigma^2_{jk_n/n} $. The proof of $ \widehat{IV}^{n} \rightarrow^{p} IV$ can be referred to \citet{BGJPS2006}.

For $A_2$, Riemann integrability implies that $A_2 \rightarrow^{p} 0 $.

\textbf{(2)} Before the proof, we note that we have the central limit theorems $\sqrt{n}(\widehat{IV}^{n} -\int_0^1\sigma_s^2ds) \rightarrow^{d} N(0,2\int_0^1\sigma_s^4ds)$ (see \citet{BGJPS2006}) and $\sqrt{k_n} (\widehat{\sigma^2}_{jk_n/n}^{n}(k_n) - \sigma_{jk_n/n}^2) \rightarrow^d N(0,2\sigma_{jk_n/n}^4)$. The last conclusion can be proved by following the consistency proof in  $\textbf{(1)}$, together with
the results $\sqrt{k_n}\mathbf{E}[|B_1|] \leq \sqrt{k_n^2\Delta_n} \rightarrow 0$ and $\mathbf{E}[(\sqrt{k_n}B_2)^2] \rightarrow 2\sigma_{jk_n/n}^4$.

Now, we are ready to give the proof of (\ref{thm1_2}). Observing that
\begin{align*}
&\mathbf{E} \Big[\Big|\sqrt{\dfrac{k_n}{2n}} \sum_{j=0}^{\lfloor n/k_n \rfloor-1 } \big(\dfrac{\sqrt{k_n}(\widehat{\sigma^2}_{jk_n/n}^{n}(k_n)-\widehat{IV}^{n})}{\sqrt{2}\widehat{IV}^{n}}\big)^2 - \sqrt{\dfrac{k_n}{2n}} \sum_{j=0}^{\lfloor n/k_n \rfloor -1  } \big(\dfrac{\sqrt{k_n}(\widehat{\sigma^2}_{jk_n/n}^{n}(k_n)-\widehat{IV}^{n})}{\sqrt{2}\sigma_0^2}\big)^2  \Big| \Big] \\
& \leq  \sqrt{\dfrac{k_n}{2n}} \sum_{j=0}^{\lfloor n/k_n \rfloor -1 }\mathbf{E} \Big[ \Big|\big(\dfrac{\sqrt{k_n}(\widehat{\sigma^2}_{jk_n/n}^{n}(k_n)-\widehat{IV}^{n})}{\sqrt{2}\widehat{IV}^{n}}\big)^2 - \big(\dfrac{\sqrt{k_n}(\widehat{\sigma^2}_{jk_n/n}^{n}(k_n)-\widehat{IV}^{n})}{\sqrt{2}\sigma_0^2}\big)^2 \Big| \Big] \\
& \leq \dfrac{C}{\sqrt{k_n}} \rightarrow 0,
\end{align*}
where the last inequality is derived by using Holder's inequality and the two central limit
theorems of the integrated volatility
and the spot volatility given above. Chebyshev's inequality implies that as $n\rightarrow \infty$, we have
\begin{equation*}
\sqrt{\dfrac{k_n}{2n}} \sum_{j=0}^{\lfloor n/k_n \rfloor-1 } \big(\dfrac{\sqrt{k_n}(\widehat{\sigma^2}_{jk_n/n}^{n}(k_n)-\widehat{IV}^{n})}{\sqrt{2}\widehat{IV}^{n}}\big)^2 - \sqrt{\dfrac{k_n}{2n}} \sum_{j=0}^{\lfloor n/k_n \rfloor-1} \big(\dfrac{\sqrt{k_n}(\widehat{\sigma^2}_{jk_n/n}^{n}(k_n)-\widehat{IV}^{n})}{\sqrt{2}\sigma_0^2}\big)^2 \rightarrow^{p} 0.
\end{equation*}
Obviously, the result ($\ref{thm1_2}$) can be obtained by showing that as $n \rightarrow \infty$,
\begin{equation*}
\begin{cases}
A_3 :=  \sqrt{\dfrac{k_n}{2n}} \sum\limits_{j=0}^{\lfloor n/k_n \rfloor -1 } \big(\dfrac{\sqrt{k_n}(\widehat{IV}^{n} -\sigma_0^2)}{\sqrt{2}\sigma_0^2}\big)^2 \rightarrow^{p} 0,\\
A_4 := \sqrt{\dfrac{k_n}{2n}} \sum\limits_{j=0}^{\lfloor n/k_n \rfloor -1} \dfrac{k_n(\widehat{\sigma^2}_{jk_n/n}^{n}(k_n)-\sigma_0^2)(\widehat{IV}^{n} -\sigma_0^2)}{\sigma_0^4} \rightarrow^{p} 0, \\
A_5 := \sqrt{\dfrac{k_n}{2n}} \sum\limits_{j=0}^{\lfloor n/k_n \rfloor -1 } \Big( \big(\dfrac{\sqrt{k_n}(\widehat{\sigma^2}_{jk_n/n}^{n}(k_n)-\sigma_0^2)}{\sqrt{2}\sigma_0^2}\big)^2 - 1\Big) \rightarrow^{d} N(0,1).
\end{cases}
\end{equation*}
For $A_3$, we have $\mathbf{E}[|A_3|] \leq C\sqrt{\dfrac{k_n}{n}} $, which follows from $\sqrt{n}(\widehat{IV}^{n} -\sigma_0^2) = O_p(1)$ and the
boundedness of $\sigma_0$.
For $A_4$, observing that $\mathbf{E}[(\sigma_0\Delta_i^nB)^2] = \sigma_0^2/n$ for $i=1,\cdots,n$, we have
\begin{align*}
\mathbf{E} [(A_4)^2] & = \dfrac{k_n}{2n} \sum_{j=0}^{\lfloor n/k_n \rfloor -1} \mathbf{E} \big[ \dfrac{k_n^2(\widehat{\sigma^2}_{jk_n/n}^{n}(k_n)-\sigma_0^2)^2(\widehat{IV}^{n} -\sigma_0^2)^2}{\sigma_0^8} \big] \\
& + \dfrac{k_n}{2n} \mathop{\sum\sum\limits}_{j_1,j_2=0, j_1\neq j_2}^{\lfloor n/k_n \rfloor -1  } \mathbf{E} \big[ \dfrac{k_n^4/n^2(\widehat{\sigma^2}_{j_1k_n/n}^{n}(k_n)-\sigma_0^2)^2(\widehat{\sigma^2}_{j_2k_n/n}^{n}(k_n)-\sigma_0^2)^2}{\sigma_0^8} \big]\\
& \leq C\dfrac{k_n}{n} \rightarrow 0,
\end{align*}
where the last inequality is derived by plugging in the conclusions of $\sqrt{n}(\widehat{IV}^{n} -\int_0^1\sigma_0^2ds) = O_p(1)$
and $\sqrt{k_n} (\widehat{\sigma^2}_{jk_n/n}^{n}(k_n) - \sigma_0^2) = O_p(1)$. For $A_5$, denote
\begin{equation*}
\xi_j = \sqrt{\dfrac{k_n}{2n}} \Big( \big(\dfrac{\sqrt{k_n}(\widehat{\sigma^2}_{jk_n/n}^{n}(k_n)-\sigma_0^2)}{\sqrt{2}\sigma_0^2}\big)^2 - 1\Big), \qquad j=0,\cdots, \lfloor n/k_n \rfloor-1.
\end{equation*}
Observing that
\begin{align*}
\mathbf{E}[\xi_j|\mathcal{F}_{jk_n/n}] & = \sqrt{\dfrac{k_n}{2n}} \dfrac{1}{2\sigma_0^4} \big(k_n\mathbf{E}[(\widehat{\sigma^2}_{jk_n/n}^{n}(k_n)-\sigma_0^2)^2|\mathcal{F}_{jk_n/n}] - 2\sigma_0^4\big) \\
& = \sqrt{\dfrac{k_n}{2n}} \dfrac{1}{2\sigma_0^4} \big( k_n(\dfrac{3\sigma_0^4}{k_n} + \dfrac{k_n(k_n-1)\sigma_0^4}{k_n^2} -\sigma_0^4) - 2\sigma_0^4 \big) \equiv 0,
\end{align*}
and it is obvious that $\xi_j$ is $\mathcal{F}_{(j+1)k_n/n}$-measurable, so that $\{ \xi_j, \mathcal{F}_{jk_n/n} \}$ is a martingale
difference array. And
\begin{align*}
\sum_{j=0}^{\lfloor n/k_n \rfloor-1 }\mathbf{E}[(\xi_j)^2 | \mathcal{F}_{jk_n/n}]& =  \dfrac{k_n}{2n} \sum_{j=0}^{\lfloor n/k_n \rfloor -1 } \big( \mathbf{E}\big[ \big(\dfrac{\sqrt{k_n}(\widehat{\sigma^2}_{jk_n/n}^{n}(k_n)-\sigma_0^2)}{\sqrt{2}\sigma_0^2}\big)^4 \big| \mathcal{F}_{jk_n/n} \big] - 1\big)\\
& = \dfrac{k_n}{2n} \sum_{j=0}^{\lfloor n/k_n \rfloor -1 } \Big( \dfrac{n^4}{4\sigma_0^8k_n^2} \big(\sum_{i=jk_n+1}^{(j+1)k_n} \mathbf{E}\big[\big( (\sigma_0\Delta_{i}^{n}B)^2 - \sigma_0^2/n\big)^4\big|\mathcal{F}_{jk_n/n}\big] \\
&+3\sum_{i_1,i_2=jk_n+1\atop i_1 \neq i_2 }^{(j+1)k_n} (\sigma_0\Delta_{i_1}^{n}B)^2 - \sigma_0^2/n\big)^2(\sigma_0\Delta_{i_2}^{n}B)^2 - \sigma_0^2/n\big)^2 \big)  - 1 \Big) \\
&= \frac{k_n}{2n} \sum_{j=0}^{\lfloor n/k_n \rfloor -1 } \Big( \dfrac{n^4}{4\sigma_0^8k_n^2}\big(k_n(60\sigma_0^8/n^4)  + 3k_n(k_n-1)(4\sigma_0^8/n^4) \big) - 1\Big) \\
& =\frac{k_n}{2n} \sum_{j=0}^{\lfloor n/k_n \rfloor-1 }( \frac{12}{k_n} +2 ) \rightarrow 1, ~~\mbox{as}~~ n \rightarrow \infty.
\end{align*}
According to the central limit theorem for martingale process stated in \citet{HH1980}, we can get $A_5 \rightarrow^{d} N(0,1)$.

\textbf{(3)} The first claim is a direct consequence of \textbf{(2)}, while the second claim follows from \textbf{(1)} and the
Portmanteau lemma.\hfill $\Box$

\textbf{Proof of Theorem \ref{thm2}:} \textbf{(1)} Under our setting, the conditions for Theorem 1 in \citet{M2009}
are satisfied, so that if $n$ is large enough, for P-almost all $\omega$, we have $\mathbf{I}_{\{(\Delta_i^n Y)^2 \leq r(1/n) \}}(\omega)= \mathbf{I}_{\{\Delta_i^n N = 0 \}}(\omega)$. Note that the stochastic integral $\int_{(i-1)/n}^{i/n}\sigma_sdB_s$ is a time changed
Brownian motion (\citet{RY2001}, Theorems 1.9 and 1.10), and by the L\'{e}vy's law for the modulus of continuity of Brownian motion's paths (\citet{KS1999}, Theorem 9.25), we have
\begin{align*}
\sup_{i\in \{1,\cdots,n \}} \dfrac{| \int_{(i-1)/n}^{i/n} \sigma_sdB_s |}{\sqrt{2\log{n}/n}}  \leq C.
\end{align*}
Together with that the total number of jumps $N_1 < C$, we have
\begin{align*}
\dfrac{n}{\log{n}}(\widehat{IV}^{n,Thr} - \widehat{IV}^{n}) & = \dfrac{n}{\log{n}}\big(\sum_{i=1}^{n} (\Delta_i^n Y)^2 \mathbf{I}_{\{(\Delta_i^n Y)^2 \leq r(1/n) \}} - \sum_{i=1}^{n} (\Delta_i^n X)^2 \big) \\
& = \dfrac{n}{\log{n}}\big(\sum_{i=1}^{n} \big(\Delta_i^n Y)^2 \mathbf{I}_{\{\Delta_i^n N = 0 \}}- \sum_{i=1}^{n} (\Delta_i^n X)^2 \big) \\
& = \dfrac{n}{\log{n}}\big(\sum_{i=1}^{n} \big(\Delta_i^n X)^2 \mathbf{I}_{\{\Delta_i^n N = 0 \}}- \sum_{i=1}^{n} (\Delta_i^n X)^2 \big) \\
& = \dfrac{n}{\log{n}} \sum_{i=1}^{n} \big(\Delta_i^n X)^2 \mathbf{I}_{\{\Delta_i^n N > 0 \}} < C,
\end{align*}
and similarly
\begin{align*}
\dfrac{n}{\log{n}}(\widehat{\sigma^2}_{jk_n/n}^{n,Thr}(k_n) - \widehat{\sigma^2}_{jk_n/n}^{n}(k_n)) < C.
\end{align*}
Obviously, we also have the corresponding versions of central limit theorems for the thresholding
estimators, namely $\sqrt{n}(\widehat{IV}^{n,Thr} -\int_0^1\sigma_s^2ds) \rightarrow^{d} N(0,2\int_0^1\sigma_s^4ds)$ and $\sqrt{k_n} (\widehat{\sigma^2}_{jk_n/n}^{n,Thr}(k_n) - \sigma_{jk_n/n}^2) \rightarrow^d N(0,2\sigma_{jk_n/n}^4)$.

According to (1) of Theorem \ref{thm1}, the result (\ref{thm2_1}) can be proved by showing
\begin{align*}
(\widehat{\sigma^2}_{jk_n/n}^{n,Thr}(k_n)-\widehat{IV}^{n,Thr} )^2 - (\widehat{\sigma^2}_{jk_n/n}^{n}(k_n)-\widehat{IV}^{n} )^2 \rightarrow^{p} 0,
\end{align*}
which naturally follows from the results mentioned above and Chebyshev's inequality.

\textbf{(2)} According to the proof of (2) of Theorem \ref{thm1}, to obtain (\ref{thm2_2}), we only need to prove
 \begin{align*}
 \sqrt{\dfrac{k_n}{2n}} \sum_{j=0}^{\lfloor n/k_n \rfloor -1 } \Big(\big(\dfrac{\sqrt{k_n}(\widehat{\sigma^2}_{jk_n/n}^{n,Thr}(k_n)-\widehat{IV}^{n,Thr})}{\sqrt{2}\widehat{IV}^{n,Thr}}\big)^2 - \big(\dfrac{\sqrt{k_n}(\widehat{\sigma^2}_{jk_n/n}^{n}(k_n)-\widehat{IV}^{n})}{\sqrt{2}\widehat{IV}^{n}}\big)^2 \Big) \rightarrow^{p} 0.
 \end{align*}
 And from the proof of (1), we have
 \begin{align*}
 &E[\Big| \sqrt{\dfrac{k_n}{2n}} \sum_{j=0}^{\lfloor n/k_n \rfloor -1 } \Big(\big(\dfrac{\sqrt{k_n}(\widehat{\sigma^2}_{jk_n/n}^{n,Thr}(k_n)-\widehat{IV}^{n,Thr})}{\sqrt{2}\widehat{IV}^{n,Thr}}\big)^2 - \big(\dfrac{\sqrt{k_n}(\widehat{\sigma^2}_{jk_n/n}^{n}(k_n)-\widehat{IV}^{n})}{\sqrt{2}\widehat{IV}^{n}}\big)^2 \Big) \Big|] \\
 &\leq C\frac{\sqrt{k_n}\log{n}}{\sqrt{n}} \rightarrow 0.
 \end{align*}
Then, Chebyshev's inequality gives us the desired result.

\textbf{(3)} The conclusion is evident from the previous proofs. \hfill $\Box$

\begin{lem}\label{lem1}
If as $n\rightarrow \infty$, $l_n\rightarrow 0$ and $n/p_n^2 \rightarrow 0$, then
\begin{align}\label{pre_cons1}
\widehat{IV}^{n,Pre}(p_n) & \rightarrow^{p}  \int_{0}^{1}\sigma_s^2ds, \\\label{pre_cons2} \widehat{\sigma^2}_{kp_nl_n/n}^{n,Pre}(p_n,l_n) & \rightarrow^{p} \sigma_{kp_nl_n/n}^2,
\end{align}
and if further $n^3/p_n^5\rightarrow 0$ and $l_nn^2/p_n^4\rightarrow 0$, then
\begin{align}\label{pre_clt1}
	\sqrt{\dfrac{n}{p_n}}(\widehat{IV}^{n,Pre}(p_n)- \int_{0}^{1}\sigma_s^2ds) & \rightarrow^{ds} N(0,\int_{0}^{1}2\sigma_s^4ds),\\\label{pre_clt2}
	\sqrt{l_n}(\widehat{\sigma^2}_{kp_nl_n/n}^{n,Pre}(p_n,l_n) - \sigma_{kp_nl_n/n}^2) & \rightarrow^{ds} N(0,2\sigma_{kp_nl_n}^4).
\end{align}

\end{lem}
\textbf{Proof:} Since the asymptotic normality of the estimators $\widehat{IV}^{n,Pre}(p_n)$ and $\widehat{\sigma^2}_{kp_nl_n/n}^{n,Pre}(p_n,l_n)$
implies their consistent convergence to $\int_{0}^{1}\sigma_s^2ds$ and $\sigma_{kp_nl_n/n}^2$, we only present the proof procedures of (\ref{pre_clt1}) and (\ref{pre_clt2}), and the relative relaxed conditions on the parameters for the consistency results can be easily seen from the following proof.

For the proof of (\ref{pre_clt1}), observing that
\begin{align*}
\sqrt{\dfrac{n}{p_n}}( \widehat{IV}^{n,Pre}(p_n) - IV )&= \sqrt{\dfrac{n}{p_n}}( \dfrac{1}{\varphi_n} \sum_{j=0}^{ \lfloor n/p_n \rfloor-1}(\overline{Z}_{jp_n}^n)^2 - IV ) \\
& =\sqrt{\dfrac{n}{p_n}} \big( \dfrac{1}{\varphi_n} \sum_{j=0}^{ \lfloor n/p_n \rfloor-1}(\overline{X}_{jp_n}^n)^2 - \dfrac{1}{\varphi_n} \sum_{j=0}^{ \lfloor n/p_n \rfloor-1}(\sigma_{jp_n/n}\overline{B}_{jp_n}^n)^2 \big) \\
&\quad + \dfrac{1}{\varphi_n} \sqrt{\dfrac{n}{p_n}} \big( \sum_{j=0}^{ \lfloor n/p_n \rfloor-1}[(\sigma_{jp_n/n}\overline{B}_{jp_n}^n)^2 -\sigma_{jp_n/n}^2\varphi_np_n/n ] \big) \\
&\quad + \sqrt{\dfrac{n}{p_n}} \big( \sum_{j=0}^{\lfloor n/p_n \rfloor-1}p_n/n\sigma_{jp_n/n}^2  -  IV \big) + \sqrt{\dfrac{n}{p_n}} \cdot O_p(\frac{n}{p_n^{2}}) \\
& := A_1' + A_2' + A_3'.
\end{align*}
The first term in $A_3'$ converges to 0 in probability, which is deduced from (6.14) in \citet{PV2009b} and its intact proof can be found in \citet{BGJPS2006}, together with the convergence $n^3/p_n^5\rightarrow 0$, we have $ A_3' \rightarrow^{p} 0$. It's obvious that the result (\ref{pre_cons1}) only requires $n/p_n^2 \rightarrow 0$. We also have $A_2' \rightarrow^{ds} N(0,\int_{0}^{1}2\sigma_s^4ds) $, which is a
special case of Lemma 3 in \citet{PV2009b}, by taking $r=2, l=0$ in $L_n(r,l)$ without the consideration of the microstructure noise. Now,
we are left to prove $A_1' \rightarrow^{p} 0$. Denote $\gamma_{j} = \sqrt{\dfrac{n}{p_n}} \dfrac{1}{\varphi_n} \big( (\overline{X}_{jp_n}^n)^2 - (\sigma_{jp_n/n}\overline{B}_{jp_n}^n)^2  \big) $, by writing
\begin{align*}
A_1' = \sum_{j=0}^{\lfloor n/p_n \rfloor-1} (\gamma_{j} - \mathbf{E} [\gamma_{j} | \mathcal{F}_{jp_n/n} ] ) +  \sum_{j=0}^{\lfloor n/p_n \rfloor-1} \mathbf{E} [\gamma_{j} | \mathcal{F}_{jp_n/n} ] := B_1' + B_2',
\end{align*}
equivalently, we only need to prove $B_1' \rightarrow^p 0$ and $B_2' \rightarrow^p 0$. Observing that $\{ \gamma_j, \mathcal{F}_{jp_n/n} \}$ is a martingale difference array, and
\begin{align*}
\mathbf{E} [(B_1')^2] = \sum_{j=0}^{\lfloor n/p_n \rfloor-1} \mathbf{E} [ (\gamma_{j} - \mathbf{E} [\gamma_{j} | \mathcal{F}_{jp_n/n} ] )^2 |\mathcal{F}_{jp_n/n} ] \leq C  \sum_{j=0}^{\lfloor n/p_n \rfloor-1} \mathbf{E} [(\gamma_{j})^2 | \mathcal{F}_{jp_n/n} ],
\end{align*}
Holder's inequality and Lemma 1 in \citet{PV2009b} yield $\sum\limits_{j=0}^{\lfloor n/p_n \rfloor-1} \mathbf{E} [(\gamma_{j})^2 | \mathcal{F}_{jp_n/n} ] \leq C/n$, thus Chebyshev's inequality implies $B_1' \rightarrow^p 0$. The proof of $B_2' \rightarrow^p 0$ can be achieved by totally
following the proof of (6.15) in \citet{PV2009b}, the only difference is regard to the varying coefficients driving the increments,
which have no effect on the convergence result. This ends the proof of (\ref{pre_clt1}).

For the proof of (\ref{pre_clt2}), similarly, we can write
\begin{align*}
\sqrt{l_n}(\widehat{\sigma^2}_{kp_nl_n/n}^{n,Pre}(p_n,l_n)-\sigma_{kp_nl_n/n}^2) & =\sqrt{l_n} ( \dfrac{n}{p_nl_n\varphi_n} \sum_{j=kl_n+1}^{ (k+1)l_n}(\overline{Z}_{jp_n}^n)^2 - \sigma_{kp_nl_n/n}^2 ) \\
& = \sqrt{l_n} ( \dfrac{n}{p_nl_n\varphi_n} \sum_{j=kl_n+1}^{ (k+1)l_n}[(\overline{X}_{jp_n}^n)^2 - (\sigma_{kp_nl_n/n}\overline{B}_{jp_n}^n)^2]) \\
&\quad + \sqrt{l_n} ( \dfrac{n}{p_nl_n\varphi_n} \sum_{j=kl_n+1}^{ (k+1)l_n} (\sigma_{kp_nl_n/n}\overline{B}_{jp_n}^n)^2 - \sigma_{kp_nl_n/n}^2)\\
&\quad + \sqrt{l_n}\cdot O_p(\frac{n}{p_n^2}) \\
& := A_4' + A_5' + A_6'.
\end{align*}
Obviously, we have $A_6' \rightarrow 0$, and we only require $n/p_n^2 \rightarrow 0$ for the consistency result (\ref{pre_cons2}).
Observing that $\mathbf{E} [ (\sigma_{kp_nl_n/n}\overline{B}_{jp_n}^n)^2  | \mathcal{F}_{kp_nl_n/n}] = \varphi_np_n/n\cdot\sigma_{kp_nl_n/n}^2$,
after some variance calculations and verifications similar to the ones in the proof of $A_2'$ above, we obtain $A_5' \rightarrow^{ds} N(0, 2\sigma_{kp_nl_n/n}^2)$. By following the proof of $A_1' \rightarrow^p 0 $ above, we also obtain $A_4' \rightarrow^p 0 $, thus we
have (\ref{pre_clt2}).\hfill $\Box$

\textbf{Proof of Theorem \ref{thm3}: } \textbf{(1)} According to the Proof of (1) of Theorem \ref{thm1}, the conclusion naturally
follows from the results $\widehat{IV}^{n,Pre}(p_n) \rightarrow^{p} IV$ and $\widehat{\sigma^2}_{kp_nl_n/n}^{n,Pre}(p_n,l_n) \rightarrow^{p} \sigma^2_{kp_nl_n/n}$.

\textbf{(2)}
Observing that
\begin{align*}
	&\mathbf{E} \Big[\Big|\sqrt{\dfrac{p_nl_n}{2n}} \sum_{k=0}^{\lfloor n/(p_nl_n) \rfloor -1} \big(\dfrac{\sqrt{l_n}(\widehat{\sigma^2}_{kp_nl_n/n}^{n,Pre}(p_n,l_n)-\widehat{IV}^{n,Pre})}{\sqrt{2}\widehat{IV}^{n,Pre}}\big)^2\\
	& \quad - \sqrt{\dfrac{p_nl_n}{2n}} \sum_{k=0}^{\lfloor n/(p_nl_n) \rfloor-1 } \big(\dfrac{\sqrt{l_n}(\widehat{\sigma^2}_{kp_nl_n/n}^{n,Pre}(p_n,l_n)-\widehat{IV}^{n,Pre})}{\sqrt{2} \sigma_0^2}\big)^2  \Big| \Big] \\
	& \leq  \sqrt{\dfrac{p_nl_n}{2n}} \sum_{k=0}^{\lfloor n/(p_nl_n) \rfloor -1 }\mathbf{E} \Big[ \Big|\big(\dfrac{\sqrt{l_n}(\widehat{\sigma^2}_{kp_nl_n/n}^{n,Pre}(p_n,l_n)-\widehat{IV}^{n,Pre})}{\sqrt{2}\widehat{IV}^{n,Pre}}\big)^2 \\
	&\quad - \big(\dfrac{\sqrt{l_n}(\widehat{\sigma^2}_{kp_nl_n/n}^{n,Pre}(p_n,l_n)-\widehat{IV}^{n,Pre})}{\sqrt{2}\sigma_0^2}\big)^2 \Big| \Big] \\
	& \leq \dfrac{C}{\sqrt{l_n}} \rightarrow 0,
\end{align*}
the last inequality is derived by using Holder's inequality and the conclusions of $\sqrt{\dfrac{n}{p_n}}(\widehat{IV}^{n,Pre} -\int_0^1\sigma_0^2ds) = O_p(1)$ and $\sqrt{l_n} (\widehat{\sigma^2}_{kp_nl_n/n}^{n,Pre}(p_n,l_n) - \sigma_0^2) = O_p(1)$. Chebyshev's inequality implies that
as $n\rightarrow \infty$, we have
\begin{align}
	 & \sqrt{\dfrac{p_nl_n}{2n}} \sum_{k=0}^{\lfloor n/(p_nl_n) \rfloor -1} \big(\dfrac{\sqrt{l_n}(\widehat{\sigma^2}_{kp_nl_n/n}^{n,Pre}(p_n,l_n)-\widehat{IV}^{n,Pre})}{\sqrt{2}\widehat{IV}^{n,Pre}}\big)^2  \notag \\
	 &  - \sqrt{\dfrac{p_nl_n}{2n}} \sum_{k=0}^{\lfloor n/(p_nl_n) \rfloor-1 } \big(\dfrac{\sqrt{l_n}(\widehat{\sigma^2}_{kp_nl_n/n}^{n,Pre}(p_n,l_n)-\widehat{IV}^{n,Pre})}{\sqrt{2} \sigma_0^2}\big)^2 \rightarrow^{p} 0.
\end{align}

Obviously, the result will hold if we can prove that as $n \rightarrow \infty$, it holds that
\begin{equation*}
	\begin{cases}
		A_1'' :=  \sqrt{\dfrac{p_nl_n}{2n}} \sum\limits_{k=0}^{\lfloor n/(p_nl_n) \rfloor-1 } \big(\dfrac{\sqrt{l_n}(\widehat{IV}^{n,Pre} -\sigma_0^2)}{\sqrt{2}\sigma_0^2}\big)^2 \rightarrow^{p} 0,\\
		A_2'' := \sqrt{\dfrac{p_nl_n}{2n}} \sum\limits_{k=0}^{\lfloor n/(p_nl_n) \rfloor -1 } \dfrac{l_n(\widehat{\sigma^2}_{kp_nl_n/n}^{n,Pre}(p_n,l_n)-\sigma_0^2)(\widehat{IV}^{n,Pre} -\sigma_0^2)}{\sigma_0^4} \rightarrow^{p} 0, \\
		A_3'' := \sqrt{\dfrac{p_nl_n}{2n}} \sum\limits_{k=0}^{\lfloor n/(p_nl_n) \rfloor -1 } \Big( \big(\dfrac{\sqrt{l_n}(\widehat{\sigma^2}_{kp_nl_n/n}^{n,Pre}-\sigma_0^2)}{\sqrt{2}\sigma_0^2}\big)^2 - 1\Big) \rightarrow^{d} N(0,1).
	\end{cases}
\end{equation*}
For $A_1''$, we have $\mathbf{E}[|A_1''|] \leq C\sqrt{\dfrac{p_nl_n}{n}} $, which follows from $\sqrt{\dfrac{n}{p_n}}(\widehat{IV}^{n}
-\sigma_0^2) = O_p(1)$ and the boundedness of $\sigma_0$, and Chebyshev's inequality implies the convergence in probability.  For
$A_2''$, since
\begin{align*}
A_2'' & =  \sqrt{\dfrac{p_nl_n}{2n}} \sum_{k=0}^{\lfloor n/(p_nl_n) \rfloor-1 } \dfrac{l_n\big(\dfrac{n}{p_nl_n\varphi_n} \sum_{j=kl_n+1}^{ (k+1)l_n}(\overline{X}_{jp_n}^n)^2-\sigma_0^2\big)\big(\dfrac{1}{\varphi_n} \sum_{j=0}^{ \lfloor n/p_n \rfloor-1}(\overline{X}_{jp_n}^n)^2 -\sigma_0^2\big)}{\sigma_0^4} \\ &\quad+O_p(\dfrac{1}{p_n}
 + \dfrac{\sqrt{nl_n}}{p_n}) \\
& :=B_2'' + O_p(\dfrac{1}{p_n} + \dfrac{\sqrt{nl_n}}{p_n}),
\end{align*}
Chebyshev's inequality implies that $A_2'' - B_2'' \rightarrow^{p} 0 $. Observing that $\mathbf{E}[\dfrac{n}{p_nl_n\varphi_n} \sum_{j=kl_n+1}^{ (k+1)l_n}(\overline{X}_{jp_n}^n)^2-\sigma_0^2] = 0 $ for $k=0,\cdots,\lfloor n/(p_nl_n) \rfloor -1$, we have
\begin{align*}
	\mathbf{E} [(B_2'')^2] & = \dfrac{p_nl_n}{2n} \sum_{k=0}^{\lfloor n/(p_nl_n) \rfloor -1} \mathbf{E} \big[ \dfrac{l_n^2(\widehat{\sigma^2}_{kp_nl_n/n}^{n,Pre}(p_n,l_n)-\sigma_0^2)^2(\widehat{IV}^{n,Pre} -\sigma_0^2)^2}{\sigma_0^8} \big] \\
	& + \dfrac{p_nl_n}{2n} \mathop{\sum\sum}_{k_1,k_2=1, k_1\neq k_2}^{\lfloor n/(p_nl_n) \rfloor -1 } \mathbf{E} \big[ \dfrac{l_n^4p_n^2/n^2(\widehat{\sigma^2}_{k_1p_nl_n/n}^{n,Pre}(p_n,l_n)-\sigma_0^2)^2(\widehat{\sigma^2}_{k_2p_nl_n/n}^{n,Pre}(p_n,l_n)-\sigma_0^2)^2}{\sigma_0^8} \big]\\
	& \leq C\dfrac{p_nl_n}{n} \rightarrow 0,
\end{align*}
the last inequality is derived by plugging in  the conclusions $\sqrt{\dfrac{n}{p_n}}(\widehat{IV}^{n,Pre} -\int_0^1\sigma_0^2ds) = O_p(1)$ and $\sqrt{l_n} (\widehat{\sigma^2}_{kp_nl_n/n}^{n,Pre}(p_n,l_n) - \sigma_0^2) = O_p(1)$, Chebyshev's inequality implies $B_2'' \rightarrow^{p} 0 $.
For $A_3''$, since
\begin{align*}
A_3'' &= \sqrt{\dfrac{p_nl_n}{2n}} \sum_{k=0}^{\lfloor n/(p_nl_n) \rfloor -1  } \Big( \big(\dfrac{\sqrt{l_n}\big(\dfrac{n}{p_nl_n\varphi_n} \sum_{j=kl_n+1}^{ (k+1)l_n}(\overline{X}_{jp_n}^n)^2-\sigma_0^2\big)}{\sqrt{2}\sigma_0^2}\big)^2 - 1\Big) + O_p(\dfrac{n}{p_n^{3/2}}) \\
& := B_3'' + O_p(\dfrac{n}{p_n^{3/2}}).
\end{align*}
Then, Chebyshev's inequality implies that $A_3'' - B_3'' \rightarrow^{p} 0$, thus we only need to prove $B_3'' \rightarrow^{d} N(0,1)$.
Denote
\begin{equation}
	\xi_k' = \sqrt{\dfrac{p_nl_n}{2n}} \Big( \big(\dfrac{\sqrt{l_n}\big(\dfrac{n}{p_nl_n\varphi_n} \sum_{j=kl_n+1}^{ (k+1)l_n}(\overline{X}_{jp_n}^n)^2-\sigma_0^2\big)}{\sqrt{2}\sigma_0^2}\big)^2 - 1\Big), \qquad k=0,\cdots, \lfloor n/(p_nl_n) \rfloor-1.
\end{equation}
By using that $\mathbf{E} [(\overline{X_i})^{2k}] = (2k-1)!!(\varphi_np_n\sigma_0^2/n)^k$ for $k=1,2,\cdots$, we have
\begin{align*}
	\mathbf{E}[\xi_k'|\mathcal{F}_{kp_nl_n/n}] & = \sqrt{\dfrac{p_nl_n}{2n}} \dfrac{1}{2\sigma_0^4} \big(l_n\mathbf{E}[\big( \dfrac{n}{p_nl_n\varphi_n} \sum_{j=kl_n+1}^{(k+1)l_n}(\overline{X}_{jp_n}^n)^2 -\sigma_0^2\big)^2|\mathcal{F}_{kp_nl_n/n}] - 2\sigma_0^4\big) \\
	& = \sqrt{\dfrac{p_nl_n}{2n}} \dfrac{1}{2\sigma_0^4} \big( l_n(\dfrac{3\sigma_0^4}{l_n} + \dfrac{l_n(l_n-1)\sigma_0^4}{l_n^2} -\sigma_0^4) - 2\sigma_0^4 \big) \equiv 0,
\end{align*}
and it is obvious that $\xi_k'$ is $\mathcal{F}_{(k+1)p_nl_n/n}$-measurable, so that $\{ \xi_k, \mathcal{F}_{kp_nl_n/n} \}$ is a martingale difference array. And
\begin{align*}
	&\sum_{k=0}^{\lfloor n/(p_nl_n) \rfloor-1 }\mathbf{E}[(\xi_k')^2 | \mathcal{F}_{kp_nl_n/n}] \\
	& =  \dfrac{p_nl_n}{2n} \sum_{k=0}^{\lfloor n/(p_nl_n) \rfloor-1 } \big( \mathbf{E}\big[ \big(\dfrac{\sqrt{l_n}\big( \dfrac{n}{p_nl_n\varphi_n} \sum_{j=kl_n+1}^{(k+1)l_n}(\overline{X}_{jp_n}^n)^2 -\sigma_0^2\big)}{\sqrt{2}\sigma_0^2}\big)^4 \big| \mathcal{F}_{kp_nl_n/n} \big] - 1\big)\\
	& = \dfrac{p_nl_n}{2n} \sum_{k=0}^{\lfloor n/(p_nl_n) \rfloor-1 } \Big( \dfrac{l_n^2}{4\sigma_0^8}\big( \mathbf{E}\big[ (\dfrac{n}{p_nl_n\varphi_n} \sum_{j=kl_n+1}^{ (k+1)l_n}(\overline{X}_{jp_n}^n)^2 )^4 \big| \mathcal{F}_{kp_nl_n/n} \big] \\
	& \qquad \qquad \qquad \qquad \qquad - \mathbf{E}\big[ 4\sigma_0^2(\dfrac{n}{p_nl_n\varphi_n} \sum_{j=kl_n+1}^{ (k+1)l_n}(\overline{X}_{jp_n}^n)^2 )^3 \big| \mathcal{F}_{kp_nl_n/n} \big] \\
	&\qquad \qquad \qquad \qquad \qquad + \mathbf{E}\big[ 6\sigma_0^4(\dfrac{n}{p_nl_n\varphi_n} \sum_{j=kl_n+1}^{ (k+1)l_n}(\overline{X}_{jp_n}^n)^2 )^2 \big| \mathcal{F}_{kp_nl_n/n} \big] \\
	& \qquad \qquad \qquad \qquad \qquad - \mathbf{E}\big[ 4\sigma_0^6(\dfrac{n}{p_nl_n\varphi_n} \sum_{j=kl_n+1}^{ (k+1)l_n}(\overline{X}_{jp_n}^n)^2 ) \big| \mathcal{F}_{kp_nl_n/n} \big] + \sigma_0^8 \big) -1\Big)\\
	& = \dfrac{p_nl_n}{2n} \sum_{k=0}^{\lfloor n/(p_nl_n) \rfloor -1} \Big( \dfrac{l_n^2}{4} \big( \dfrac{l_n^4+12l_n^3+44l_n^2+48l_n}{l_n^4} -\dfrac{4(l_n^3+6l_n^2+8l_n)}{l_n^3} + \dfrac{6(l_n^2+2l_n)}{l_n^2} -3 \big) \Big) \\
	& =\frac{p_nl_n}{2n} \sum_{k=0}^{\lfloor n/(p_nl_n) \rfloor -1}( \frac{12}{l_n} +2 ) \rightarrow 1.
\end{align*}
According to the central limit theorem for martingale process in \citet{HH1980}, the results above implies that $B_3'' \rightarrow^{d} N(0,1)$, which ends the proof.

\textbf{(3)} The conclusion is obvious from the established results. \hfill $\Box$

\bigskip
\bibliographystyle{model2-names}
\bibliography{liu}

\end{document}